\documentstyle[11pt]{article}
\input rotate.tex
\input psfig.tex

\begin{document}

\begin{center}
{\LARGE Polarization Effects in Superdeformed Nuclei}\\
\vspace{1cm}

{\large Lennart B. Karlsson, Ingemar Ragnarsson and Sven {\AA}berg} \\
\vspace{0.5cm}

Department of Mathematical Physics, Lund Institute of Technology, \\
P.O.\ Box 118, S-221 00 Lund, Sweden
\end{center}
\vspace{1cm}

\noindent
{\bf Abstract:} A detailed theoretical investigation of polarization
effects in superdeformed nuclei is performed. In the pure harmonic
oscillator potential it is shown that when one particle (or hole) with
the mass single-particle quadrupole moment $q_{\nu}$ is added to a
superdeformed core, the change of the electric quadrupole moment can be
parameterized as $q_{eff}=e(bq_{\nu}+a)$, and analytical expressions
are derived for the two parameters, $a$ and $b$. Simple numerical
expressions for $q_{eff}(q_{\nu})$ are obtained in the more realistic
modified oscillator model. It is also shown that quadrupole moments of
nuclei with up to 10 particles removed from the superdeformed core of
$^{152}$Dy can be well described by simply subtracting effective
quadrupole moments of the active single-particle states from the
quadrupole moment of the core. Tools are given for estimating the 
quadrupole moment for possible configurations in the superdeformed 
$A \sim 150$-region.

\noindent
{\it PACS:} 21.60.-n, 21.10.Re, 21.10.Ky

\noindent
{\it Keywords:} Polarization, quadrupole moment, harmonic oscillator,
Nilsson-Strutinsky, superdeformation.

\section{Introduction}

Superdeformed rotational bands were first identified \cite{Twi86} from
transition energies only. At an early stage, crude measurements of
(electric) quadrupole moments were performed \cite{Ben87} which were
an important evidence for the large deformation. Recently, it has
become possible to measure relative quadrupole moments with a much
higher precision \cite{Sav96,Nis97,Hac98}. Since the total quadrupole
moment depends sensitively on the specific orbitals that are occupied,
these quadrupole moments have become an important tool to verify
configurations of different bands. It is the aim of the present paper
to study in detail how the quadrupole moment depends on occupied
single-particle states in superdeformed nuclei.

Adding a particle with a single-particle quadrupole moment, $q_{\nu}$,
to a core causes a change in deformation due to polarization effects.
The size of the deformation change depends on the relative
deformations, or quadrupole moments, of the core and the added
particle. The deformation change can be translated to a change in the
electric quadrupole moment and we shall refer to this change as the
effective quadrupole moment $q_{eff}$. We will consider two different
ways to calculate $q_{eff}$. One (microscopic) way is to calculate the
total quadrupole moment from a sum of single-particle quadrupole
moments at the minimum-energy deformations for the configurations
before and after the particle is added. The other (macroscopic) way is
to calculate the quadrupole moment from a homogeneously charged body
with the appropriate deformation and volume. The change of the
electric quadrupole moment when one particle/hole is added then
constitutes the effective electric quadrupole moment.

Simple models \cite{Mot58} suggest that a near-spherical $Z=N$ nucleus
changes its microscopic electric quadrupole moment with about
$q_{\nu}/2$ due to polarization, and by an additional $q_{\nu}$ if the
added particle is a proton. Generalizing to a deformed nucleus we
shall show that, in the pure harmonic oscillator (HO) model, $q_{eff}$
can be written as
\begin{equation}
\left( q_{eff} \right)_{p,n}=e \left(b_{p,n}(\varepsilon) \cdot q_{\nu} 
+ a_{p,n}(\varepsilon) \right).
\end{equation}
The parameters $b_{p,n}$ and $a_{p,n}$ depend on deformation,
$\varepsilon$, and are different for protons $(p)$ and neutrons $(n)$.
The relation (1) will be derived in section~2 and, for the microscopic
as well as the macroscopic methods, analytic expressions will be
given for $b_{p,n}$ and $a_{p,n}$ at superdeformation but also at
other closed-shell configurations corresponding to different
deformations.

When comparing experimental and calculated quadrupole moments, the
modified oscillator (MO) has often been used. In refs.\
\cite{Sav96,Nis97}, the deformation for different superdeformed
configurations was calculated using the Nilsson-Strutinsky cranking
method and the quadrupole moment was then obtained in the
macroscopic way. The agreement between experiment and theory was
generally found to be quite good. Furthermore, these quadrupole
moments come close to those obtained in relativistic mean field
calculations \cite{Afa96}.

In section~3 the quadrupole moment is studied using the MO potential. The
polarization effects from adding one particle (or hole) to the yrast
superdeformed $^{152}$Dy configuration are studied in subsection 3.1
utilizing the macroscopic approach to calculate the quadrupole
moment. The corresponding results from the microscopic approach are
presented in subsection~3.3. In both models we obtain simple relations
corresponding to eq.~(1), but with some modifications. The possible
reasons for these modifications (cranking, hexadecapole deformation,
Strutinsky renormalization, the $l^2$-term or the $\vec{l}\cdot
\vec{s}$-term in the potential) are analysed.

In the models used here, superdeformed bands are understood from
single-particle motion in a rotating deformed potential. Then, it
seems natural to ask if physical quantities can be described by adding
effective contributions from particles in different orbitals. The
relations found in ref.~\cite{Ben88} between high-$N_{osc}$
configurations in superdeformed bands and the ${\cal J}^{(2)}$ moments
of inertia were based on the additivity of single-particle ${\cal
J}^{(2)}$ contributions. An attempt to test the additivity of
experimental ${\cal J}^{(2)}$ moment of inertia was performed in
ref.~\cite{Del89}. For the single-particle angular momentum,
additivity of alignment was first tried in ref.~\cite{Rag91}, then
further tested in refs.~\cite{Rag93,preprint} and found to work
well. Similarly, it was concluded that specific orbitals lead to
well-defined deformation changes corresponding to an additivity for
deformations \cite{Rag93}.

Deformations can be translated into quadrupole moments whose
additivity were tested in selfconsistent Skyrme-Hartree-Fock
calculations by Satu{\l}a {\it et al}.\ \cite{Sat96}. They extracted
effective quadrupole moments from least square fits over a large
number of configurations in the region of nuclei with $Z=64-67,
N=84-87$, and found that the quadrupole moments of these
configurations could be well described by summing the contributions
from the orbitals involved. In the present paper, somewhat similar
studies are described in subsections~3.2 and 3.4 using the more
simplistic cranked Nilsson-Strutinsky approach. However, our studies
cover the whole $A=152$ superdeformed region from $^{142}$Sm to
$^{152}$Dy where a limited number of low-lying configurations are
investigated in detail. Furthermore, comparisons are made between
extrapolations based on calculated effective and bare single-particle
quadrupole moments, respectively. Indeed, based on our analytical
calculations in the HO, we get a microscopic understanding why
additivity works so well also in realistic nuclear models. In
subsection~3.5 we compare with experimental data. Finally, a short
summary is given in section~4.

\section{Quadrupole moments in the pure oscillator.}

The polarization effect on a deformed core by one particle is studied
in the pure HO potential. There exist two equivalent methods
to calculate this effect (see e.g.\ p.~510 in ref.~\cite{BM2}):
{\em Either} as a renormalization of the quadrupole operator due to the
coupling of the single-particle excitations to the 
(isoscalar) giant quadrupole resonance (treated in RPA),
{\em or} by considering the new, self-consistent deformation that results
when one particle is added to a core. In the present paper, only the latter 
method is used. 

In subsection~2.1 some useful definitions are given, and an effort is 
made to calculate the polarization using double-stretched coordinates.
In subsection~2.2 non-stretched coordinates are used and
explicit expressions are derived for the polarization effect on an
axially symmetric $Z=N$ core. In subsection~2.3 the formulae are
generalized to a core with $Z \neq N$.

\subsection{Double-stretched quadrupole moments for one kind of particles}

Consider a HO potential 
\begin{equation}
V_{osc}=\frac 12m\left( \omega _x^2x^2+\omega _y^2y^2+\omega _z^2z^2\
\right). 
\end{equation}
A specific orbital $\left| \nu \right\rangle =\left| n_xn_yn_z\right\rangle$ 
is described by the number of quanta $n_i$ in the three Cartesian
directions.  The single-particle energies are given by $e_\nu $ and
the single-particle mass quadrupole moments $q_\nu $ are calculated as
\begin{equation}
q_\nu =\langle \nu \mid 2z^2-x^2-y^2\mid \nu \rangle . 
\end{equation}
In the potential, the equipotential surfaces are ellipsoidal with the axes
proportional to $1/\omega _i$. 
We will only consider axial symmetric solutions corresponding to $\omega
_x=\omega _y=\omega _{\bot }$. Elongation is then described by the standard
parameter $\varepsilon $ \cite{Nil55},
\begin{eqnarray}
\omega _z &=&\omega _0\left( 1-\frac{2\varepsilon }3\right)   \nonumber \\
\omega _x &=&\omega _y=\omega _0\left( 1+\frac \varepsilon 3\right). 
\end{eqnarray}
Volume conservation corresponds to 
\begin{equation}
\omega _x\omega _y\omega _z=\left( \stackrel{o}\omega _{0}\right) ^3 ,
\end{equation}
where the parameter $\stackrel{o}\omega _{0}$ is determined from the
radius of the core. We now transform the physical coordinates
$\left(x,y,z\right) $ to dimensionless coordinates and furthermore
introduce the stretched coordinate system \cite{Nil55,Lar72}, 
\begin{eqnarray}
x^{\prime } &=&\xi =\sqrt{\frac{\hbar \omega _x}m}x,  \nonumber \\
y^{\prime } &=&\eta =\sqrt{\frac{\hbar \omega _y}m}y, 
\label{str} \\
z^{\prime } &=&\zeta =\sqrt{\frac{\hbar \omega _z}m}z,  \nonumber
\end{eqnarray}
corresponding to the system where the eigensolutions separate in
the three Cartesian directions. The single-particle quadrupole moment
becomes 
\begin{equation}
q_\nu =\frac{\hbar ^2}m\frac{\stackrel{o}\omega _{0}}{\omega _0}\left( \frac{%
2\left( n_z+1/2\right) }{1-2\varepsilon /3}-\frac{n_x+1/2}{1+\varepsilon /3}-%
\frac{n_y+1/2}{1+\varepsilon /3}\right). 
\end{equation}

For a system of $Z$ particles (protons) in the potential, we calculate
the total energy $E$ as the sum of the single-particle energies under the
constraint of volume conservation. The energy can be written as 
\begin{equation}
E=\hbar\omega _x\Sigma_x+\hbar\omega _y\Sigma_y+\hbar\omega _z\Sigma_z,
\end{equation}
where the $\Sigma ^{\prime }s$ measure the total number of quanta in the
different directions, 
\begin{equation}
\Sigma _i=\sum_{\nu =1}^{Z} \left( n_i+\frac 12\right)  _\nu \quad ; 
\quad i=x,y,z.
\label{Sigmai}
\end{equation}
The selfconsistent deformation, is obtained by
minimizing the total energy, in the ($\varepsilon,\gamma$) deformation space.
The energy minimization is equivalent to a self-consistency between the 
potential and the matter distribution \cite{BM2}, namely that the ratio 
$\left\langle x^2\right\rangle :$ 
$\left\langle y^2\right\rangle :$ 
$\left\langle z^2\right\rangle $ is the same for the two distributions. 
This can be expressed as
\begin{equation}
\Sigma _x\omega _x=\Sigma _y\omega _y=\Sigma _z\omega _z.
\end{equation}

From now on we will assume axial symmetry corresponding to an equal
number of quanta in the two perpendicular directions, 
\begin{equation}
\Sigma _x = \Sigma_y \equiv \frac{1}{2}\Sigma_{\perp}. 
\label{Sigma}
\end{equation}
The axial symmetry corresponds to $\gamma=0$ while $\varepsilon$
is obtained as (see e.g.\ refs.\
\cite{Cer79,Nil95})
\begin{equation}
\varepsilon =\frac{3\left( 2\Sigma_z-\Sigma_{\perp} \right) }
{4\Sigma_z+\Sigma_{\perp} }.
\label{eps1}
\end{equation}

The total microscopic electric quadrupole moment $Q^{mic}$ is calculated as
the sum of the single-particle quadrupole moments
\begin{equation}
Q^{mic}=e\sum_{\nu = 1}^{Z} q_\nu =\frac{\hbar ^2 e}m\cdot\frac{\stackrel{o}
\omega _{0}}{\omega _0}\left( \frac{2\Sigma _z}
{1-2\varepsilon /3}-\frac{\Sigma_{\perp} }
{1+\varepsilon /3}\right). 
\end{equation}
The question which specifically interests us now is how $Q^{mic}$ is changed
if {\em one} particle with a single-particle (mass) 
quadrupole moment $q_\nu $ is added to the core.
Adding one particle to a spherically symmetrical HO potential 
($\Sigma _x=\Sigma_y=\Sigma _z; Q^{mic}=0$), it is well-known \cite{Mot58} 
that the total quadrupole 
moment becomes $Q^{mic}=2eq_\nu $.

Starting from an arbitrary axially symmetric deformation of the
deformed HO, defined by the total number of quanta in
the different directions, $\Sigma _z$ and $\Sigma_{\perp},$ and the number of
particles $Z$, a similar relation is found, however only when
expressed in the double-stretched coordinates \cite{Sak89},
\begin{equation}
x^{\prime \prime }=\frac{\hbar \omega _{\bot }}mx,\qquad y^{\prime \prime }=%
\frac{\hbar \omega _{\bot }}my,\qquad z^{\prime \prime }=\frac{\hbar \omega
_z}{m}z.
\end{equation}
We define the single-particle quadrupole moment in these coordinates, 
\begin{equation}
q_\nu ^{\prime \prime }=\langle \nu \mid 2\left( z^{\prime \prime }\right)
^2-\left( x^{\prime \prime }\right) ^2-(y^{\prime \prime })^2\mid \nu
\rangle. 
\end{equation}
It is then straightforward to show that at the self-consistent deformation $%
\varepsilon _0$ defined by eq.~(\ref{eps1}), 
\begin{equation}
Q^{\prime \prime }=e\sum_\nu q_\nu ^{\prime \prime }=0,
\end{equation}
i.e.\ the matter distribution is always `spherically symmetric' in the
double-stretched system.

When adding a particle with a double-stretched quadrupole moment $%
q_{\nu}^{\prime \prime }$, we find a total double-stretched quadrupole
moment,
\begin{equation}
Q^{\prime \prime }\left( \varepsilon _0\right) =2eq_\nu ^{\prime \prime },
\label{Qtotds}
\end{equation}
where $Q^{\prime \prime }$ is calculated at the `new' self-consistent
deformation, but where $\varepsilon _0$ indicates that the
double-stretched coordinates are now defined with $\omega _{\bot }$
and $\omega _z$ corresponding to the original deformation,
$\varepsilon _0$, which is different from the new deformation obtained
with the added particle. This is thus a generalization of the formula
for spherical shape.

However, it turns out that eq.~(\ref{Qtotds}) is not too useful when
calculating how the physical quadrupole moment $Q$ is influenced by
the addition of a particle. It is straightforward to find linear
relations between $Q$ and $\langle r^2\rangle $ in the two systems
(and also in the single-stretched system, eq.~(\ref{str})), but the
somewhat complicated form of these relations, and the fact that also
$\langle r^2 \rangle $ in the double-stretched system must then be
analyzed, make us conclude that it is more straightforward to work
directly with the formulae in the physical (non-stretched)
coordinates.

\subsection{Quadrupole moments with a $Z=N$ core.}

Consider a system of $Z$ protons (with charge e) and $N$
neutrons (with no charge) in their respective HO potential with the
total number of quanta (see eqs.\ (\ref{Sigmai}, \ref{Sigma}))
described by ($\Sigma_{zp},\Sigma_{\perp p} $) and 
($\Sigma _{zn},\Sigma_{\perp n}$),
respectively. The protons and neutrons are coupled in the standard way
used in the MO \cite{Nil69}, namely that protons and neutrons have the
same radius. For nuclear radii to be reproduced, this results in
frequencies varying according to
\begin{equation}
\hbar \stackrel{o}\omega _{0}
=\frac D{A^{1/3}}\left( 1\mp \frac{N-Z}{A}\right)^{1/3}
\approx \frac D{A^{1/3}}\left( 1\mp \frac{N-Z}{3A}\right), 
\label{hbarom}
\end{equation}
where the standard value of $D$ is 41~MeV corresponding to
$r_{o}=1.2$~fm. The rightmost expression has become standard in MO
calculations and will be used by us except when discussing explicit
$Z$ and $N$ dependences. We then note that
\begin{eqnarray}
\frac {1}{A^{1/3}}\left( 1- \frac{N-Z}{A}\right)^{1/3}
&=&\frac{(2Z)^{1/3}}{A^{2/3}} \nonumber \\ 
\frac {1}{A^{1/3}}\left( 1+ \frac{N-Z}{A}\right)^{1/3}
&=&\frac{(2N)^{1/3}}{A^{2/3}}.
\label{ZNhbarom}
\end{eqnarray}
For the isolated systems of protons or neutrons, we can use
eq.~(\ref{eps1}) to calculate the deformation of an arbitrary state,
and thus the deformation change, $\delta \varepsilon $, if a particle
is added. For simplicity we start by analysing a nucleus with an equal
number of protons and neutrons in the core. By minimizing the total energy,
i.e.\ the sum of the single-particle energies,
of the $Z=N$ proton-neutron system the equilibrium deformation is obtained as
\begin{equation}
\varepsilon =\frac 12\left(\frac{3\left(2\Sigma _{zp}-\Sigma_{\perp p}\right)}
{4\Sigma _{zp}+\Sigma _{\perp p}}+
\frac{3\left( 2\Sigma _{zn}-\Sigma _{\perp n}\right) }
{4\Sigma _{zn}+\Sigma _{\perp n}}\right). 
\label{epspn}
\end{equation}
This expression will be derived for the general $Z \neq N$ HO system 
in subsect.~2.3 below, see eq.~(\ref{gendef}). Equation (\ref{epspn}) 
gives the reasonable result that to lowest order in
$\delta \varepsilon$, the addition of either a proton or a neutron 
leads to a deformation change $\delta \varepsilon /2$, for the combined 
system. This is a simple model which can be studied analytically leading to
closed formulae which should be helpful when considering shape changes
and polarization effects in the more realistic MO model. 

Let us first note that the formulae in the previous subsection can easily
be generalized to the $Z=N$ system. Thus, if a particle with a mass
quadrupole moment $q_{\nu}$ is added to a spherical system, because of
its polarization is `equally divided' between protons and neutrons,
the total charge quadrupole moment will increase by $0.5eq_{\nu}$ if
the added particle is a neutron and by $(1+0.5)eq_{\nu}$ for a proton.
Furthermore, eq.~(\ref{Qtotds}) can be generalized in an analogous way
for the $Z=N$ system.

For a deformed core we assume that the added particle has $n_z$
quanta in the axial direction and $n_x$, $n_y$ quanta in the
perpendicular directions. In an analogous way to the capital $\Sigma
$'s above (cf.\ eq.~(\ref{Sigmai})), describing the total number of
quanta, we then introduce
\begin{eqnarray}
\sigma _z &=&n_z+1/2  \nonumber \\
\sigma_{\perp}  &=&n_x+n_y+1.
\end{eqnarray}
The single-particle quadrupole moment can now be written as,
\begin{equation}
q_\nu =\frac{\hbar ^2}{mD}\cdot \frac{\stackrel{o}\omega _{0}}{\omega _0}%
\cdot \frac{A^{1/3}}{1\mp \frac{N-Z}{3A}}\left( \frac{2\sigma _z}{1-\frac{%
2\varepsilon }3}-\frac{\sigma_{\perp} }{1+\frac \varepsilon 3}\right), 
\end{equation}
and the total {\em microscopic} electric quadrupole moment as, 
\begin{equation}
Q^{mic}=e\sum\limits_{\nu \in prot}q_\nu =
\frac{\hbar ^2 e}{mD}\cdot \frac{\stackrel{o}\omega _{0}}{\omega _0}\cdot 
\frac{A^{1/3}}{1- \frac{N-Z}{3A}}
\left( \frac{2\Sigma _{zp}}{1-\frac{2\varepsilon }3}-
\frac{\Sigma_{\perp p} }{1+\frac\varepsilon 3}\right), 
\label{Qmic}
\end{equation}
where the volume conservation factor is 
\begin{equation}
\frac{\stackrel{o}\omega _{0}}{\omega _0}=\left( 1+\frac \varepsilon
3\right) ^{2/3}\left( 1-\frac{2\varepsilon }3\right) ^{1/3}.
\end{equation}

We can also define a {\em macroscopic} electric quadrupole moment
$Q^{mac}$ calculated from an isotropic charge distribution with the
total charge equal to $Ze.$ For such a spheroid with the symmetry axis
$b$ and the perpendicular axis $a$, the quadrupole moment is given as
$(2/5)Ze(b^2-a^2)$.  Then with a radius parameter $r_0,$ i.e.\ a volume
$(4/3)\pi r_0^3A,$ and a quadrupole deformation calculated from the
$\Sigma $'s according to eq.~(\ref{epspn}), the macroscopic electric
quadrupole moment becomes,
\begin{equation}
Q^{mac}=\frac 15ZeA^{2/3}r_0^2\left( \frac{\stackrel{o}\omega _{0}}{\omega _0}%
\right) ^2\left( \frac 2{\left( 1-\frac{2\varepsilon }3\right) ^2}-\frac
2{\left( 1+\frac \varepsilon 3\right) ^2}\right).
\label{Qmac} 
\end{equation}
From eqs.\ (\ref{Qmac},\ref{Qmic}), using eq.~(\ref{ZNhbarom}), we
see that both $Q^{mac}$ and $Q^{mic}$ are proportional to $ZA^{2/3}$
as $\Sigma_{\perp p}$ and $\Sigma _{zp}$ increase as $Z^{4/3}$. Since
$\varepsilon $ can be expressed in the $\Sigma $'s, $Q^{mac}$ and
$Q^{mic}$ can be considered as functions of the independent variables
$\Sigma_{\perp} $ and $\Sigma _z$ for protons and neutrons, respectively,
together with the number of protons $Z$ and neutrons $N.$

We now determine how $Q^{mic}$ (eq.~(\ref{Qmic})) and $Q^{mac}$
(eq.~(\ref{Qmac})) are changed from the addition of one proton by
differentiating with respect to $Z$, $\Sigma _{\perp p}$ and $\Sigma _{zp}$,
and from the addition of one neutron by differentiating with respect
to $N$, $\Sigma _{\perp n}$ and $\Sigma _{zn}$. The final expressions can be
simplified quite a lot if instead of the original $\Sigma _{\perp}$ and
$\Sigma _z$ (which are the same for neutrons and protons), we
introduce the axis-ratio $k$ between the z-axis and the perpendicular
axes, which can be written as
\begin{equation}
k=\frac{\Sigma _z}{\frac{1}{2}\Sigma_{\perp}},
\end{equation}
and use $k$ and $\Sigma_{\perp}$ as independent variables. For example, the
volume conservation factor then takes the form
\begin{equation}
\frac{\stackrel{o}\omega _{0}}{\omega _0}=
\frac{3 k^{2/3}}{ 2k+1}. 
\end{equation}
We will use $A_0$ for the number of particles in the reference nucleus which
thus has $A_0/2$ protons and $A_0/2$ neutrons. Then, with $k$ measuring the
deformation, $q_\nu $ takes the simple form
\begin{equation}
q_\nu =\frac{\hbar ^2}{Dm}k^{-1/3}A_0^{1/3}(2kn_z+k+n_z-1-N_{osc}),\label{qny}
\end{equation}
where instead of $\sigma _z$ and $\sigma_{\perp} ,$ we have used 
$n_z$ and $N_{osc}=n_x + n_y + n_z$ to characterize the particle.

We will refer to the change in $Q$ ($Q^{mic}$ or $Q^{mac}$)
when a particle is added as $q_{eff},$
which can generally be expressed in $A_0,$ $\Sigma_{\perp} $ and $k,$ in 
addition to 
$n_z$ and $N_{osc}$, describing the properties of the added particle. These
expressions are not very complicated but become even simpler if we put 
$k=2$, corresponding to an axis ratio of $2:1$, i.e.\ a superdeformed shape: 
\begin{eqnarray}
\left( q_{eff}^{mac}\right) _p &=&\left( A_0/2\right) ^{2/3}r_0^2 e
\left( 1.6+\frac{A_0}{\Sigma_{\perp} }\left( 1.2n_z-0.3-0.6N_{osc}\right) 
\right)  \nonumber \\
\left( q_{eff}^{mac}\right) _n &=&\left( A_0/2\right) ^{2/3}r_0^2 e
\left( 0.4+\frac{A_0}{\Sigma_{\perp} }\left( 1.2n_z-0.3-0.6N_{osc}\right) 
\right)  \nonumber \\
\left( q_{eff}^{mic}\right) _p &=&\frac{\hbar ^2e}{Dm}\left( A_0/2\right)
^{1/3}\left( 8n_z+0.25-2.5N_{osc}\right)  \label{Qeff} \\
\left( q_{eff}^{mic}\right) _n &=&\frac{\hbar ^2e}{Dm}\left( A_0/2\right)
^{1/3}\left(\frac{2 \Sigma_{\perp}}{A_0}+ 3n_z-0.75-1.5N_{osc}
\right).  \nonumber
\end{eqnarray}
Note that these formulae are general for a superdeformed $Z=N$ system
in the sense that we do not require 
that the $\left( A_0/2\right)$ lowest orbitals are filled; the only 
requirement is that $\Sigma_z=\Sigma_{\perp}$ for the core.

An interesting physical situation corresponds to the filling of the
orbitals below the $2:1$ gaps of the HO. It is then important to note
that only for every second gap, the deformation calculated
(eq.~(\ref{eps1})) from the minimum of the sum of the single-particle
energies corresponds to a $2:1$ ratio of the nuclear axes
($\varepsilon =0.6),$ i.e.\ it is only for these {\it selfconsistent gaps} 
that $\Sigma _z=\Sigma_{\perp}.$ These are the $Z=N=g=4,16,40,80,140,\dots
$ gaps, where we have introduced $g$ for the particle number at these
selfconsistent gaps. It is clear (see eqs.\ (\ref{Qeff}) and (\ref{qny})) that
if for one of these gaps, $q_{eff}$ is plotted vs. $q_\nu $ with
varying $N_{osc}$ for fixed $n_z$, or with varying $n_z$ for fixed
$N_{osc}$, we obtain straight lines. If both $n_z$ and $N_{osc}$ are
varied, the relation is not as evident, however.

Of main physical interest are the orbitals close to the Fermi surface,
and we note that the orbitals which are degenerate at $2:1$ shape have
a specific relation between $n_z$ and $N_{osc},$ i.e.\ when $N_{osc}$
decreases by one, $n_z$ decreases by 2. Thus, for each of these
bunches of degenerate orbitals, we will again get straight lines when
$q_{eff}$ is plotted vs.  $q_\nu .$ The relation between $N_{osc}$ and
$n_z$ is used to reduce the set of variables, and all the needed
information about the number of quanta can be expressed by $N_{sh}$,
which counts the number of shells at a deformation where the ratio
$\omega_z: \omega _{\bot }$ can be expressed by small integers.

Before we write down the simplified expressions, we note that the
relations at $2:1$ deformation can easily be generalized to a $k:1$
deformation where $k$ is a small integer number. Selfconsistent
 gaps are then formed
for particle numbers $Z=N=g= 2k, 8k, 20k, 40k, 70k, \dots$, i.e.\ if
the orbitals below these gaps are occupied, then 
$\Sigma_z = k\frac{1}{2}\Sigma_{\perp}$.
Furthermore, for degenerate orbitals at $k:1$ deformation, if $N_{sh}$
differs by $(k-1)$, then $n_z$ differs by $k$. For the selfconsistent gaps, 
$g$ and $\Sigma _{\perp}$ can be expressed in $N_{sh}$:
\[
g=\frac {1}{3k^2}\left( N_{sh}+1\right) \left( N_{sh}+1+k\right) \left(
N_{sh}+1+2k\right) 
\]
\begin{equation}
\Sigma_{\perp} =\frac {1}{6k^3}\left( N_{sh}+1\right) \left( N_{sh}+1+k\right)
^2\left( N_{sh}+1+2k\right), 
\end{equation}
i.e.\
\begin{equation}
\frac {g}{\Sigma_{\perp}} =\frac{2k}{\left( N_{sh}+1+k\right) }.
\end{equation}
Using these relations together with eq.~(\ref{Qeff}) in its general form 
for an arbitrary $k$-value, we obtain $q_{eff}$ as functions in 
$q_\nu $, eq.~(\ref{qny}), for selfconsistent gaps at spherical ($k=1$), 
superdeformed ($k=2$), hyperdeformed 
($k=3$), etc.\ shape,
\begin{eqnarray}
\label{qefftot}
\left( q_{eff}^{mac}\right) _p &=&\left( \frac{2}{k} \right) ^{1/3}
\frac{2g^{4/3}}{5\Sigma _{\perp}}
\frac{r_0^2 e}{\hbar^2/Dm}q_\nu - \frac{r_0^2 e}{15}
\left(\frac{2g}{k}\right)^{2/3}(k^2-1)
\left(0 \pm \frac {4}{ N_{sh}+k+1 } \right) \nonumber \\
\left( q_{eff}^{mac}\right) _n &=&\left( \frac{2}{k} \right) ^{1/3}
\frac{2g^{4/3}}{5\Sigma _{\perp}}
\frac{r_0^2 e}{\hbar^2/Dm}q_\nu -  \frac{r_0^2 e}{15}
\left(\frac{2g}{k}\right)^{2/3}(k^2-1)
\left(6 \pm \frac {4}{ N_{sh}+k+1 }\right) \nonumber \\
\left( q_{eff}^{mic}\right) _p &=&1.5eq_\nu \quad -\frac{\hbar ^2 e}{Dm}%
\left(\frac{2g}{k}\right)^{1/3} \frac{k^2-1}{3k}
\left( N_{sh}+k+1 \pm \frac{1}{2}\right) \\
\left( q_{eff}^{mic}\right) _n &=&0.5eq_\nu \quad -\frac{\hbar ^2 e}{Dm}%
\left(\frac{2g}{k}\right)^{1/3} \frac{k^2-1}{3k} \frac{1}{2}
\left( N_{sh}+k+1 \pm 1\right). \nonumber
\end{eqnarray}
All these functions are seen to be on the linear form 
\begin{equation}
q_{eff} = e(b q_\nu  +a)
\label{ba}   
\end{equation}
with
\begin{equation}
a = a_0 \pm \Delta 
\label{adelta}  
\end{equation}
where $a_0$ is the average of $q_{eff}$ for a particle and a hole with
$q_\nu =0,$ and $\Delta $ is the small deviation from this average due
to particle or hole nature of the orbital, i.e.\ a particle added to
one of the degenerate orbitals (with $N_{osc}=N_{sh}+1$) just above
the gap, or removed from one of the degenerate orbitals (with
$N_{osc}=N_{sh}$) just below the gap. For {\em prolate shape $a_0$ as
well as $\Delta$ in eq.~(\ref{adelta}) are negative} and the plus
(minus) sign in this equation corresponds to a particle
(hole).

\begin{figure}[t]
\vspace{-2.5cm}
\psfig{figure=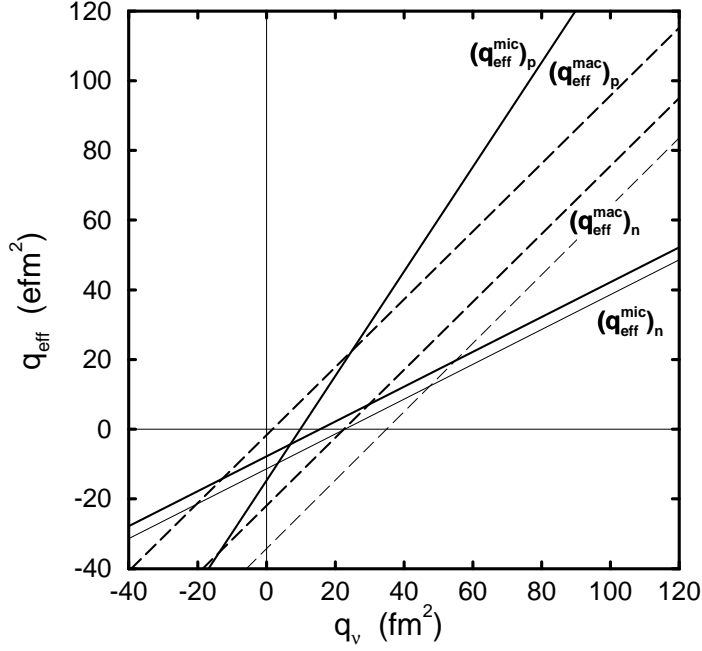,height=8cm}
\vspace{2.5cm}
\caption{Effective electric quadrupole moments, $q_{eff}$, 
versus the single-particle mass quadrupole moment, $q_\nu$, at
superdeformation, derived in the HO in the case when
one particle is added to an orbital just above the gap. Solid and
dashed lines are used for microscopic and macroscopic $q_{eff}$,
respectively. Thick lines are used for the $g=40$ gap and thin lines
for the $g=80$ gap (only drawn for neutrons). Very similar behaviour
appears (not shown) when one particle is removed from an orbital just
below the gap, i.e.\ $\Delta$ in eq.~(\ref{adelta}) is of minor importance.}
\label{fig1}
\end{figure}

Equations (\ref{qefftot}) are valid for arbitrary k values. They are 
illustrated for the superdeformed ($k=2$) $g=40$
gap ($N_{sh}=5$) in Fig.~1.  There we have primarily chosen to put the
particle in an orbital above the gap with ($N_{osc}, n_z$) = (6,6),
(5,4), etc. As $\left( q_{eff}^{mac}\right) _p$ has $a_0$ $=$ $0$
(which holds for all selfconsistent gaps) $\Delta ^{mac}$ is directly seen as
the deviation from 0 for $q_{\nu} =0$. All $a$-values are negative,
i.e.\ the$q_{eff}$-values are all negative for $q_{\nu}=0$.  This is
easily understood from the fact that the quadrupole moment of an added
particle to a superdeformed core must be sufficiently large to induce
an increased deformation.

The slopes for the microscopic $%
q_{eff}$ are the same as for the spherical case, i.e.\ $(b^{mic})_p=1.5$ and
$(b^{mic})_n=0.5$. The slopes, i.e.\ the $b$-values, in
the macroscopic case depend on both $k$ and $g$ but,
as illustrated in Fig.~2, in a way so that they 
\begin{figure}[tb]
\vspace{-2.5cm}
\psfig{figure=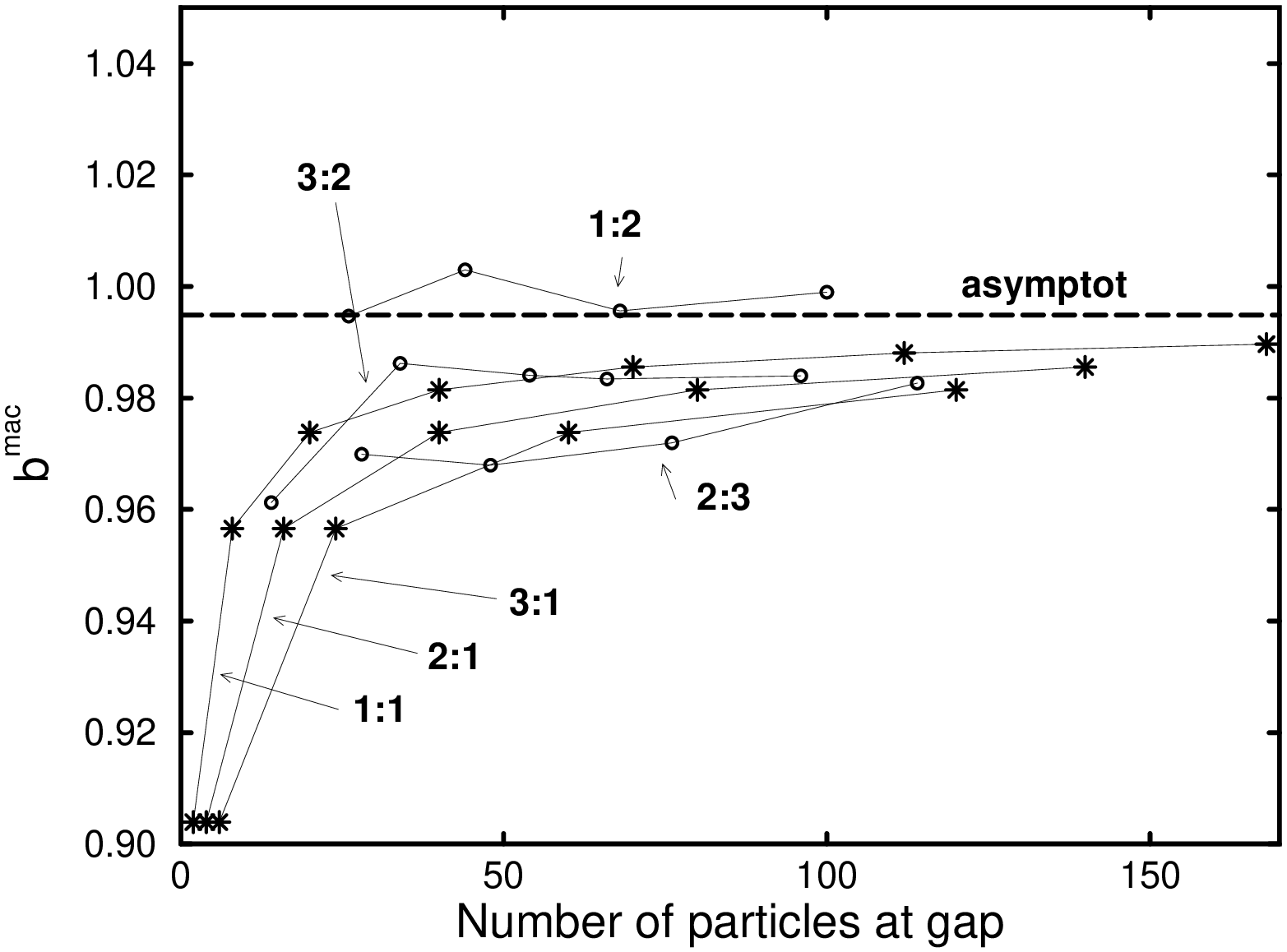,height=8cm}
\vspace{2.5cm}
\caption{Values of $b^{mac}$ (i.e.\ the slope of the dashed lines in Fig.~1.) 
as functions of number of particles of each kind ($Z=N$) in the
nucleus.  The stars connected by solid lines are the 1st, 2nd, 3rd
... selfconsistent gaps for each of the deformations: spherical ($1:1$),
superdeformed ($2:1$) and hyperdeformed ($3:1$).  One can see that the
values converge to an asymptotic value close to one (see text for
details), and that it requires the same number of selfconsistent gaps (i.e.\
more particles at larger deformation) to get the asymptotic value. The
circles show the values of the slopes for deformations ($1:2$, $2:3$
and $3:2$) where there are no selfconsistent gaps.  Note that all values are
quite close to the value at the asymptotic limit and that, as
explained in the text, the differences from this value can largely be
understood from the choice of the parameter $\hbar$$\stackrel{o}\omega_{0}$.}
\label{fig2}
\end{figure}
are almost constant and close to one
for all gaps (except for the very lowest ones). The $b^{mac}$-values
are the same e.g.\ for $g=40$ at $2:1$ as for $g=20$ at $1:1$ 
(and $g=60$ at $3:1$). 
These are the third lowest selfconsistent gaps at each deformation.  

Using eq.~(\ref{hbarom}) with the standard value of $D=41$ MeV an
asymptotic value of 0.995 is found for $b^{mac}$ when N$_{sh}
\longrightarrow \infty$.  It turns out, however, that the deviation
from 1 is simply related to the value chosen for $D$. This value is
generally determined so that $R_{rms}=1.2 A^{1/3}$ fm , see
e.g.\ \cite{Nil95}, leading to $D=41.2$ MeV with one digit higher
accuracy. Using this value of $D$ instead gives $b^{mac} = 1.000$ in the
limit $N_{sh} \longrightarrow \infty$.  Furthermore, if when deducing
$D$, we do not go to the asymptotic limit but require that $R_{rms}=1.2
A^{1/3}$ fm for each HO configuration, $\hbar$$\stackrel{o}\omega _{0}$
will depend on $g/k$ in such a way that $b^{mac} \equiv 1$ for all
selfconsistent gaps at $k:1$ deformation.  Consequently, in these cases Fig.~2
just shows the effect of a constant $\hbar$$\stackrel{o}\omega _{0}$
for all particle numbers at each deformation.  However, as $D=41$ MeV is
generally used in MO calculations independent of
particle number, we think it is interesting to illustrate the effect
which this $A$-dependence leads to for the polarization of an added
particle. In particular this means that, in microscopic calculations,
$R_{rms}$ will be larger than $1.2 A^{1/3}$ fm for light nuclei.
                                         
At spherical shape $q_{eff}=0$ for $q_\nu =0$, i.e.\ $a=0$, as seen in
eq.~(\ref{qefftot}) ($k=1$). For prolate shape $\left( a^{mic}\right)_{p,n} $
and $(a^{mac})_n$ are negative.  In Fig.~3
\begin{figure}[htb]
\vspace{-2.5cm}
\psfig{figure=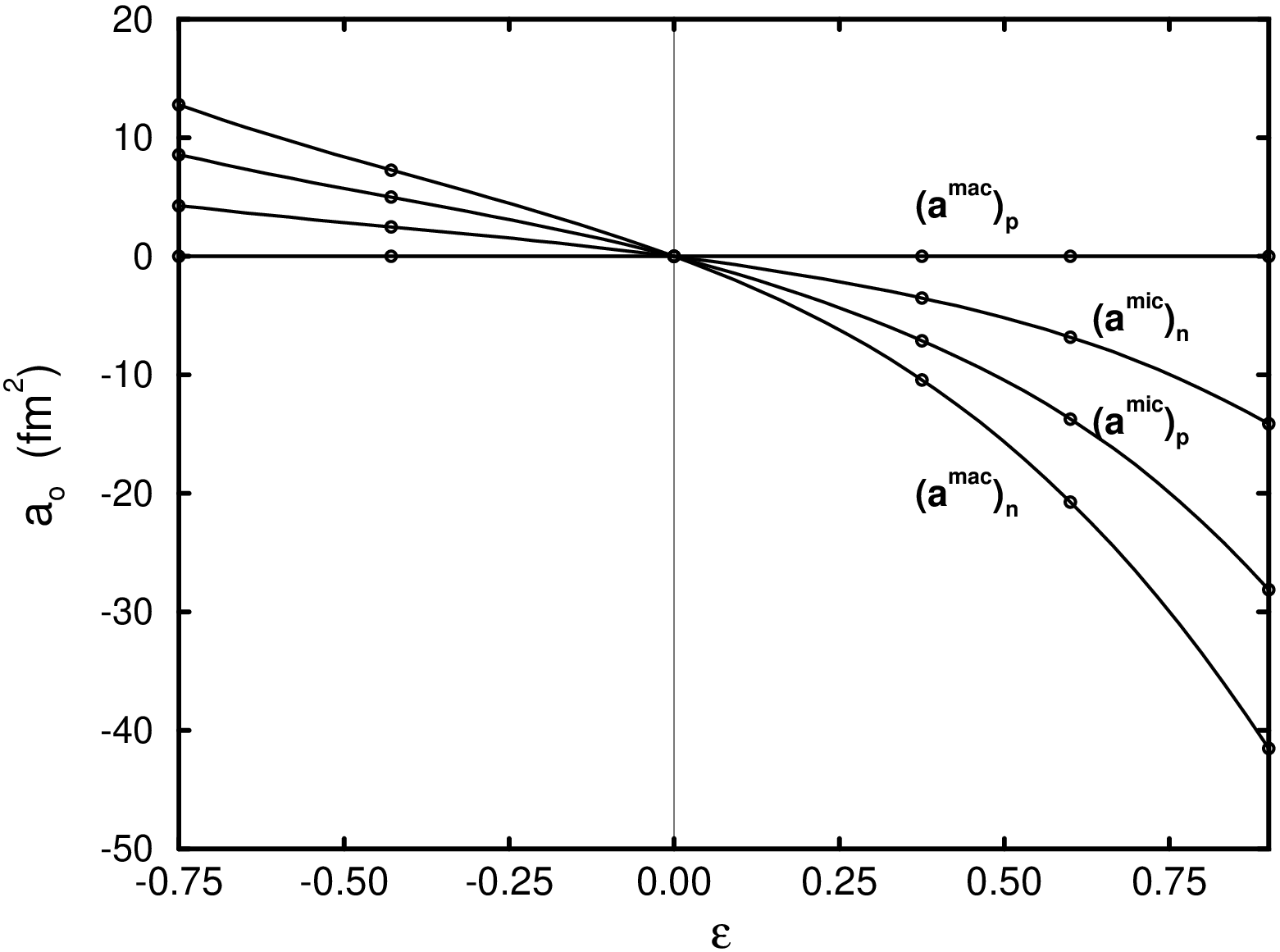,height=8cm}
\vspace{2.5cm}
\caption{Dependence of $a^{mac}$ and $a^{mic}$ on deformation $\varepsilon$.
The values are given for $g=40$ but, as the values scale with the
number of particles in the same way as the single-particle quadrupole
moment, $a/q_{\nu}$ is (asymptotically) independent of the number of
particles. The different behaviour for particles and holes ($\Delta$
in eq.~(\ref{adelta})) has been ignored.}
\label{fig3}
\end{figure}
is shown how $a$ varies with quadrupole deformation for a fixed number
of particles. For $1:1$, $2:1$ and $3:1$ deformation we have used the
selfconsistent gaps while for $1:2$, $2:3$ and $3:2$ deformation we have used
the gaps whose deformation at minimal total energy is closest to the
`correct' deformation. For all deformations we have fitted a curve as
a function of particle number (which becomes almost linear) to get the
values at $g=40$.

In the high-$N_{sh}$ limit and for selfconsistent gaps we get the asymptotic 
relations defining the macroscopic effective quadrupole moments, 
\begin{eqnarray}
\left( b^{mac}\right) _p &=&\left( b^{mac}\right) _n = 0.995 \nonumber \\
\left( a^{mac}\right) _p &=&-\left( 0 \pm 0.423\frac{k^2-1}{k^{4/3}} g^{1/3} 
\right) \label{lennart1}\\
\left( a^{mac}\right) _n &=&-\left( 0.914\frac{k^2-1}{k^{2/3}} g^{2/3}
\pm 0.423\frac{k^2-1}{k^{4/3}} g^{1/3} \right), \nonumber \\
\nonumber 
\end{eqnarray}
and the microscopic effective quadrupole moments,
\begin{eqnarray}
\left( b^{mic}\right) _p &=&1.5 \nonumber  \\
\left( b^{mic}\right) _n &=&0.5 \nonumber \\
\left( a^{mic}\right) _p &=&
-\left(0.613\frac{k^2-1}{k^{2/3}} g^{2/3} 
\pm 0.212\frac{k^2-1}{k^{4/3}} g^{1/3} \right) \label{lennart2} \\
\left( a^{mic}\right) _n &=&
-\left(0.306\frac{k^2-1}{k^{2/3}} g^{2/3} 
\pm 0.212\frac{k^2-1}{k^{4/3}} g^{1/3} \right). \nonumber \\
\nonumber
\end{eqnarray}
It is clearly seen that $a_0$ (eq.~(\ref{adelta})) increases with
$g^{2/3}$ ($\propto A^{2/3}$) which is the same growth rate as for
$q_{\nu}$, and therefore the relative importance of $a_0$ is
independent of particle-number. The polarization difference between
particles and holes, $\Delta$, on the other hand gets less important
the heavier the nucleus is as it only increases as $g^{1/3}$.

Note that although the functions for macroscopic and microscopic
$q_{eff}$ are quite different, the sum of the $q_{eff}$ values when
adding one neutron and one proton is much more alike. This means that
the total electric quadrupole moment for a nucleus with $Z \approx N$ is
approximately the same independently of if the microscopic or the
macroscopic formula is used.

The different behaviour of macroscopic and microscopic effective
quadrupole moments have the same origin in the deformed case as the
spherical case, and is caused by the two basically different
assumptions applied. In the {\em macroscopic} approach, the
protons as well as the neutrons are assumed to have a constant matter
distribution inside the potential, i.e. also the matter
distribution for protons and neutrons have identical deformations.
Consequently, for a $Z=N$ nucleus the addition of a proton or
neutron has the same polarization effect, except for a constant factor
caused by the added charge, i.e.\ same $b$- but different
$a$-parameters, as seen in eq.~(\ref{qefftot}) for a deformation
corresponding to a $k:1$ prolate shape. Independent of
deformation, we asymptotically obtain $\left( b^{mac}\right) _p
=\left( b^{mac}\right) _n \approx 1$, see eq.~(\ref{lennart1}) (see
also Fig.~1 for the k=2 case).

To analyze the {\em microscopic} approach, we start from identical
proton and neutron configurations corresponding to an equilibrium
deformation $\varepsilon_o$. Then if a proton is added, the isolated
proton system will get a new equilibrium deformation which we will
refer to as $\varepsilon_o +  \delta \varepsilon$.  The deformation
of the combined system is then $\varepsilon_o + \delta
\varepsilon/2$. The equilibrium deformation of the isolated neutron
system is unchanged, $\varepsilon_o$. The electric quadrupole moment
is then calculated at the (non-selfconsistent) deformation
$\varepsilon_o +\delta \varepsilon /2$. From eq, (12), it is easy to
find out that in a pure proton system, half of the change in the
quadrupole moment will be caused by the changed value of
$\varepsilon$, and half of the change by the change in the
$\Sigma$'s. In the present case with a proton added to a $Z=N$ system,
the change in the $\Sigma$'s for the protons is the same as when a
proton is added to a pure proton system while the change in
deformation for the total system (the polarization) is half of that
for the pure proton system. This gives $\left( b^{mic}\right) _p =
1.5$.  If instead a neutron is added, there is no contribution from
the change in the $\Sigma$'s as they refer to the
proton-configuration, but the deformation of the total system will
still be $\varepsilon_o + \delta \varepsilon/2$, and the contribution
from this change in deformation will thus be the same. This gives
$\left( b^{mic}\right) _n = 0.5$, see eq.~(\ref{lennart2}) (see also
Fig.~1 for the k=2 case).

The difference between the macroscopic and microscopic
polarizations are thus caused by the lack of selfconsistency when the
isolated proton and neutron systems have different equilibrium
deformations. Then, in the macroscopic approach, it is assumed that the
proton and neutron {\em matter distributions} fully adopt the
common deformation while they only do it 'halfway' in the microscopic
approach. With the assumption that the proton-proton, neutron-neutron
and proton-neutron attractions are the same, it seems that the
microscopic approach should come close to a fully selfconsistent
treatment. However, a stronger attraction between unlike particles
would correspond to a polarization in the direction suggested by the
macroscopic approach.
 
In subsections 3.1 and 3.3 we calculate $q_{eff}$ in
both methods for the modified oscillator potential, and in subsection
3.5 we compare the two methods with experimental data. First we will
however study the polarization effects in the HO with a $Z \neq N$
core.

\subsection{Quadrupole moments with a $Z \neq N$ core.}
How will the results from the previous subsection change when the number
of protons is far from the same as the number of neutrons? We will not
derive all the formulae again but just investigate how different
quantities depend on $Z$ and $N$. To analyse quadrupole moments, the
correct equilibrium deformation of the proton-neutron system has to be
obtained. This is done by minimizing the total energy
$E(\varepsilon,Z,N)$. The energy of the proton system or neutron
system can be written as
\begin{equation}
E_{i}=3 \hbar \stackrel{o}\omega _{0i} 
\left(\frac{1}{4}\Sigma_{\perp j}^2\Sigma_{zj}\right)^{1/3} \hspace{1cm}; 
\hspace{1cm}j=p,n.
\end{equation} 
We know that $\hbar$$\stackrel{o}\omega _{0}$ follows eq.~(\ref{ZNhbarom}),
and $\Sigma_{\perp}$ and $\Sigma_z$ are proportional to $Z^{4/3}$ and
$N^{4/3}$ for protons and neutrons, respectively. Altogether the
energy becomes
\begin{eqnarray}
E_{p} &\propto& \frac{Z^{5/3}}{ A^{2/3}} \nonumber \\
E_{n} &\propto& \frac{N^{5/3}}{ A^{2/3}}
\label{prop}
\end{eqnarray}
for the two different systems. In a region around their respective
equilibrium deformation the proton and the neutron energies can be
well approximated with parabolas. The total energy, which is just the
sum of the two energies, can therefore be written
\begin{equation}
E_{tot}=E_{0p}+C_{p}(\varepsilon-\varepsilon_{p})^{2}+
E_{0n}+C_{n}(\varepsilon-\varepsilon_{n})^{2},
\end{equation}
where the equilibrium deformation of each subsystem is
\begin{equation}
\varepsilon_{i}=\frac{3(2\Sigma_{zj}-\Sigma_{\perp j})}
{4\Sigma_{zj}+\Sigma_{\perp j}} 
\hspace{1cm}; \hspace{1cm}j=p,n
\end{equation}
and the energies at the minima ($E_{0p}$ and $E_{0n}$) and the
stiffness parameters ($C_{p}$ and $C_{n}$), all with proportionality
according to eq.~(\ref{prop}), determine each parabola. By minimizing the
total energy, $E_{tot}$, the equilibrium deformation of the total system
is obtained as
\begin{equation}
\varepsilon_{0}=\frac{C_{p}\varepsilon_{p}+C_{n}\varepsilon_{n}}{C_{p}+C_{n}}=
\frac{Z^{5/3}\varepsilon_{p}+N^{5/3}\varepsilon_{n}}{Z^{5/3}+N^{5/3}}.
\label{gendef}
\end{equation}

The $b$ values describe the part of $q_{eff}$ which depends on single
particle properties of the added particle (eq.~(\ref{ba})). They enter
via the change $\delta \varepsilon$ in equilibrium deformation, and
for protons in the microscopic case by a direct contribution of
1~$q_{\nu}$. The change in total quadrupole moment caused by the
deformation change is to first order
\begin{eqnarray}
\delta Q_{def} \propto Z A^{2/3} \delta \varepsilon,
\end{eqnarray}
while the single-particle quadrupole moment is
\begin{eqnarray}
q_{\nu p}  &\propto& \frac {A^{2/3}}{Z^{1/3}}f(\sigma_{zp},\sigma_{\perp p})  
\nonumber \\
q_{\nu n}  &\propto& \frac {A^{2/3}}{N^{1/3}}f(\sigma_{zn},\sigma_{\perp n}),
\end{eqnarray}
where $f(\sigma_{zp},\sigma_{\perp p})$ and $f(\sigma_{zn},\sigma_{\perp n})$ 
are  independent of $Z$ and $N$.
In total we get the expressions 
\begin{eqnarray}
\left(\frac{\delta Q_{def}}{q_{\nu}}\right)_{p} &\propto& \frac{Z A^{2/3} 
\frac{Z^{5/3}}{Z^{5/3}+N^{5/3}}
\frac{3(2\sigma_{zp}-\sigma_{\perp p})}{4\Sigma_{zp}+\Sigma_{\perp p}} }
{A^{2/3}Z^{-1/3}f(\sigma_{zp},\sigma_{\perp p})} =
\frac{Z^{5/3}}{Z^{5/3}+N^{5/3}}F(\sigma_{zp},\sigma_{\perp p}) \nonumber \\
\left(\frac{\delta Q_{def}}{q_{\nu}}\right)_{n} &\propto& \frac{Z A^{2/3} 
\frac{2N^{5/3}}{Z^{5/3}+N^{5/3}}
\frac{3(2\sigma_{zn}-\sigma_{\perp n})}{4\Sigma_{zn}+\Sigma_{\perp n}} }
{A^{2/3}N^{-1/3}f(\sigma_{zn},\sigma_{\perp n})  } =
\frac{Z}{N} \frac{N^{5/3}}{Z^{5/3}+N^{5/3}}F(\sigma_{zn},\sigma_{\perp n}).
\end{eqnarray}
where $F(\sigma_{zp},\sigma_{\perp p})$ and $F(\sigma_{zn},\sigma_{\perp n})$ 
are independent of $Z$ and $N$. The previously derived $b$-values
(eqs.~(\ref{lennart1},\ref{lennart2})) can now be generalized to any
proton-neutron system
\begin{eqnarray}
\left( b^{mac}(Z,N)\right) _p &=& \frac{2Z^{5/3}}{Z^{5/3}+N^{5/3}} 
\left( b^{mac}(Z_{0}=N_{0})\right) _p \nonumber  \\
\left( b^{mac}(Z,N)\right) _n &=& \frac{Z}{N} \frac{2N^{5/3}}{Z^{5/3}+N^{5/3}}
\left( b^{mac}(Z_{0}=N_{0})\right) _n   \nonumber \\
\left( b^{mic}(Z,N)\right) _p &=& 1+\frac{2Z^{5/3}}{Z^{5/3}+N^{5/3}} 
\left(\left( b^{mic}(Z_{0}=N_{0})\right) _p-1\right) \label{scaling}\\
\left( b^{mic}(Z,N)\right) _n &=& \frac{Z}{N} \frac{2N^{5/3}}{Z^{5/3}+N^{5/3}}
\left( b^{mic}(Z_{0}=N_{0})\right) _n \nonumber
\end{eqnarray}
keeping in mind that the $Z=N$ $b$-values were only deduced for selfconsistent
gaps, but should be approximately valid for all $Z$- and $N$-values.
For the nucleus $^{152}$Dy the predictions from the HO is
$b^{mac}_p=0.78$, $b^{mac}_n=0.93$, $b^{mic}_p=1.39$, and
$b^{mic}_n=0.47$.

\section{Quadrupole moments at superdeformation in the cranked MO
potential.}

We will now continue to analyze polarization effects on quadrupole
moments in the cranked MO model. Starting from the superdeformed
$^{152}$Dy configuration, we shall study how the quadrupole moment is
affected by adding or removing protons or neutrons in specific
orbitals. Contrary to the previous section, all calculations are
performed at $I \approx 40 \hbar$. It has previously been concluded
\cite{Rag80} that at superdeformation, the general properties of 
the single-particle orbitals are almost unaffected by spin up to the
highest observed values $I=60-70 \hbar$. Furthermore, our present study
shows that the polarizing properties in the HO at superdeformation are
essentially the same at $I=0 \hbar$ and at $I=40 \hbar$. In
subsections~3.1 and 3.2 we study the macroscopic quadrupole moments,
which have been used in recent comparisons with experiment and seem to
work quite well
\cite{Sav96,Nis97}, while similar calculations using the microscopic
quadrupole moments are carried out in subsections~3.3 and 3.4. We
perform a complete calculation in the cranked MO potential (parameters
from ref.~\cite{Haa93}) with Strutinsky renormalization to get the
proper equilibrium deformation, $\varepsilon $ and $\varepsilon_4$, of
each configuration. The formalism of ref.~\cite{Ben85} is used which
makes it straightforward to study specific configurations defined by
the number of particles of signature $\alpha = 1/2$ and $\alpha =
-1/2$, respectively, in the different $N_{rot}$ shells of the rotating
HO basis.  The macroscopic electric quadrupole moment, $Q^{mac}$, is
obtained from an integration over the volume described by the nuclear
potential and the microscopic electric quadrupole moment, $Q^{mic}$, by
adding the contributions $q_{\nu}$ from the occupied proton orbitals
at the proper deformation. One-particle polarization effects are
investigated in subsections~3.1 and 3.3 while the additivity of
several particles is studied in subsections~3.2 and 3.4. In
subsection~3.5 the two theoretical methods are compared and confronted
with experiment.

\subsection{Macroscopic polarization effects of one-particle states}

In the present study, we first calculate the macroscopic quadrupole
moments $Q^{mac}$ in neighbouring nuclei which only differ by one
particle or one hole from the yrast superdeformed configuration in
$^{152}$Dy. From these $Q^{mac}$-values effective electric quadrupole
moments, $q_{eff}$, are obtained for different orbitals. For the
orbitals around the $Z=66$ and $N=86$ gaps, $q_{eff}$ is plotted vs.\
the mass single-particle quadrupole moment, $q_{\nu}$, in Fig.~4 (see
also Table~4 below).
\begin{figure}[bt]
\psfig{figure=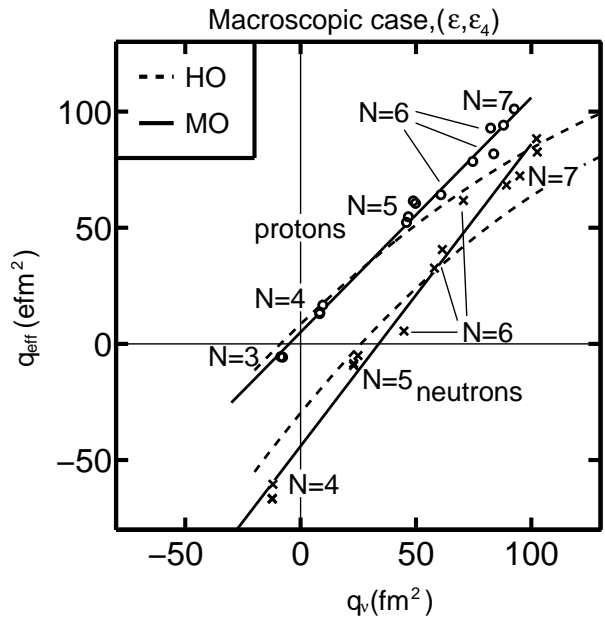,height=8cm}
\caption{Changes in the total macroscopic quadrupole moment, $q_{eff}$, 
when a particle is added to or removed from the superdeformed core
plotted vs. the single-particle mass quadrupole moment, $q_{\nu}$. The
circles and crosses are obtained from calculations in the
Strutinsky-renormalized cranked MO with simultaneous
energy-minimization in $\varepsilon$ and $\varepsilon_4$ direction
using $^{152}$Dy as a core (values are given in Table~4). The quantum
number $N \equiv N_{rot}$ is indicated for each state. The solid
lines, whose equations are given in the text, are linear least square
fits to the respective points.  The dashed lines are the quadratic
least square fits for the HO potential (with $\varepsilon_4$ deformation
included) with the ($Z=60, N=80$) core, as defined in Fig.~5.}
\label{fig4}
\end{figure}
It is evident that these values define approximate straight lines. By a
least square fit we get the relation 
\begin{equation}
q_{eff}=(1.02q_{\nu}+5.4)\; e{\rm fm}^2
\label{macp}
\end{equation}
for protons and 
\begin{equation}
q_{eff}=(1.30q_{\nu}-43.7)\; e{\rm fm}^2
\label{macn}
\end{equation}
for neutrons. This should be compared with $b^{mac}_p=0.78$ and
$b^{mac}_n=0.93$ obtained from eqs.~(\ref{lennart1},\ref{scaling}).
It is evident that there are important differences. 

Let us analyze the differences between the $Z=N$ HO and the MO results
for $^{152}$Dy by introducing, in the numerical calculations, the
different terms one at the time. Starting with the $Z=N$ HO we get the
expected modifications when combining the HO superdeformed gaps
$Z=60$ and $N=80$: Due to the neutron excess the average slope gets
notably smaller than in the $Z=N$ case and neutrons have a larger slope
than protons, all in agreement with eq.~(\ref{scaling}). Cranking the
system and performing a Strutinsky renormalization 
(i.e. including macroscopic liquid drop energy) only have a minor
effect on the slopes, see dashed lines in Fig.~5.
\begin{figure}[htb]
\psfig{figure=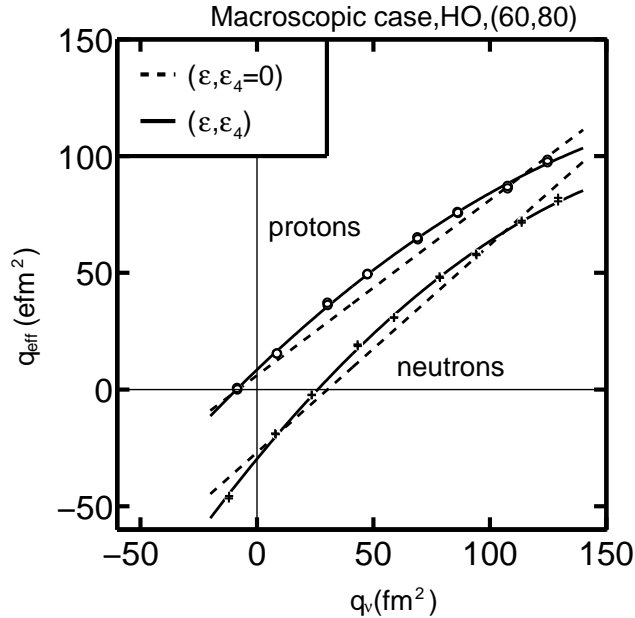,height=8cm}
\caption{Changes in the total macroscopic quadrupole moment, $q_{eff}$, when
a particle is added to or removed from the $Z=60$ and $N=80$ core
plotted vs. the single-particle mass quadrupole moment, $q_{\nu}$.
The curves are least square fits to the calculations in
Strutinsky-renormalized cranked HO calculations. The dashed lines show the
relations when the energy is minimized only in $\varepsilon$ deformation
with $\varepsilon_4=0$. The solid lines (corresponding proton and neutron data
marked with circles and crosses, respectively) show the relations when
the energy is simultaneously minimized in the $\varepsilon$ and
$\varepsilon_4$ directions.}
\label{fig5}
\end{figure}
In the standard HO only quadrupole deformation $\varepsilon$ is used,
while in the MO the energy is minimized in the quadrupole and
hexadecapole deformation plane. In order to test the importance of the
hexadecapole $\varepsilon_4$-deformation it is now included in the
(cranked and Strutinsky-renormalized) HO, see the solid lines in
Fig.~5. Due to the hexadecapole $\varepsilon_4$-deformation
there is now a curvature in the relation $q_{eff}=q_{eff}(q_{\nu})$,
but the best linear fit is very close to the one obtained for
quadrupole deformation only.

When changing to the MO potential by including the $\vec{l}\cdot
\vec{s}$ and $l^2$-terms, there are large changes, see Fig.~4. The large 
differences between the HO and the MO are understood by considering
the stiffness of the core in the two models. Around the equilibrium
quadrupole deformation $\varepsilon _0$ of the core, its energy can be
expressed as
\begin{equation}
E_{core}=E_0(\varepsilon _0)+C(\varepsilon-\varepsilon _0)^2,
\end{equation}
and the energy of the added particle as
\begin{equation}
e_{part}=e_0(\varepsilon _0)+K(\varepsilon-\varepsilon _0).
\end{equation}
The change in deformation of the total system, due to the added
particle, is therefore $K/2C$. The stiffness $C$ of the energy surface
is larger for the HO than for the MO (see below), and this directly
gives a larger deformation change in the MO for the same single
particle quadrupole moment (same $K$). Consequently, in the MO the
total quadrupole moment will increase more for high-$N_{osc}$ orbitals
with a positive deformation change and decrease more for low-$N_{osc}$
orbitals with negative deformation change. The slope of
$q_{eff}(q_{\nu})$ will therefore be considerably larger in the MO
than in the HO.

The net result of the change from the HO with a $Z=N$ core to the
cranked MO potential with the $Z=66$, $N=86$ core is thus an increase
in the slope of the $q_{eff}$ vs.\ $q_{\nu}$ relation for neutrons,
but almost no change in the proton slope.  Furthermore, the relations
still become approximately linear also in the MO case.

One might ask why the polarization effects are larger in the MO than
in the HO. To investigate this, we carried out calculations in the HO
both with the HO closed shell configuration ($Z=60, N=80$), and for the
configuration corresponding to the yrast superdeformed band in
$^{152}$Dy. It turned out that, when corrected for the difference in
mass, the different configurations had very similar stiffness. 
Calculations were then carried through for the $^{152}$Dy configuration 
with only the $l^2$-term or the $\vec{l}\cdot \vec{s}$-term included, 
indicating that the $l^2$-term is responsible for approximately 60 \% 
and the $\vec{l}\cdot \vec{s}$-term  for 40 \%
of the change in stiffness.

\subsection{Additivity of macroscopic effective quadrupole moments}

It is now interesting to investigate if the $q_{eff}$ values are additive,
i.e.\ if quadrupole moments of several-particle several-hole configurations
relative to the superdeformed $^{152}$Dy yrast state (with the $Q^{mac}$ 
value 1893 $e$fm$^2$) can be calculated from the formula
\begin{equation}
Q_{est}=Q\left( ^{152}Dy_{yrast}\right)
+\sum\limits_{particle}q_{eff}-\sum\limits_{hole}q_{eff}.  \label{Qadd}
\label{Qest}
\end{equation}
The additivity will first be tested for excited states in $^{152}$Dy 
where the number of excited particles are the same as the number of excited 
holes. We shall then test the additivity in other superdeformed nuclei all 
the way down to $^{142}$Sm. 

Starting with $^{152}$Dy, we give in Table~1 calculated deformations
and quadrupole moments for a few $n$-particle $n$-hole configurations
with rather low excitation energy. The quadrupole moments are
calculated both by exact integration and by use of eq.~(\ref{Qadd})
with  $q_{eff}$ taken from the 1-hole or 1-particle configurations
given in Fig.~4. The agreement between these two methods is good with
a typical difference of 2 $e$fm$^2$.

\begin{table}[htb]
\caption{Calculated deformations and macroscopic quadrupole moments for SD
configurations of $^{152}$Dy. The two values of the quadrupole moment are
obtained from a numerical integration ($Q_{exact}$) and by adding and
subtracting $q_{eff}$ values to the yrast quadrupole moment ($Q_{est}$). 
The values of $q_{eff}$ are obtained from 1-particle or 1-hole
configurations relative to the yrast state, see Fig.~4 and Table~4 below.}
\label{tab1}\hspace{-0.1cm}
\par
\begin{tabular}{|ccccc|}
\hline
Configurations of $^{152}$Dy. & $Q_{exact}$ & $Q_{est}$ & $\varepsilon 
$ & $\varepsilon _4$ \\ 
& ($e$fm$^2$) & ($e$fm$^2$) &  &  \\ \hline
yrast & 1893.2 &  & 0.5820 & 0.0166 \\ 
&  &  &  &  \\ 
$\pi ([651]3/2^-)^{-1}([413]5/2^-) $ & 1826.9 & 1828 & 0.5716 & 0.0211 \\ 
&  &  &  &  \\ 
$\pi ([301]1/2^-)^{-1}([532]3/2^-) $ & 1960.6 & 1960 & 0.6012 & 0.0282 \\ 
&  &  &  &  \\ 
$\nu ([770]1/2^+)^{-1}([402]5/2^+) $ & 1749.1 & 1750 & 0.5519 & 0.0135 \\ 
&  &  &  &  \\ 
$\pi ([651]3/2^-)^{-1}([651]3/2^+)^{-1} $ &  &  &  &  \\ 
$([413]5/2^-)([413]5/2^+) $ & 1765.2 & 1766 & 0.5608 & 0.0241 \\ 
&  &  &  &  \\ 
$\pi ([651]3/2^-)^{-1}([651]3/2^+)^{-1} $ &  &  &  &  \\ 
$([532]3/2^-)([413]5/2^-) $ & 1809.1 & 1811 & 0.5706 & 0.0259 \\
&  &  &  &  \\ 
$\pi ([651]3/2^-)^{-1}([413]5/2^-) $ &  &  &  &  \\ 
$\nu ([770]1/2^+)^{-1}([402]5/2^-)$ & 1686.4 & 1685 & 0.5415 & 0.0182 \\  
&  &  &  &  \\ 
$\pi ([651]3/2^-)^{-1}([651]3/2^+)^{-1}$ &  &  &  &  \\ 
$([532]3/2^-)([413]5/2^+)$ & 1726.9 & 1723 & 0.5601 & 0.0388 \\ 
$\nu ([770]1/2^+)^{-1}([521]3/2^-)$ &  &  &  &  \\ \hline
\end{tabular}
\end{table}


Considering the good agreement between the `exact' and `estimated'
macroscopic quadrupole moments for $^{152}$Dy, one could ask if 
similar methods could be used to estimate the quadrupole moment 
for other SD nuclei around $^{152}$Dy. Furthermore, considering the 
approximately linear relation between $q_{\nu}$ and $q_{eff}$ discussed 
above, it is interesting to investigate if these quadrupole
moments can be obtained from a knowledge of $Q$ for the reference nucleus
and $q_{\nu}$ (but not $q_{eff}$) for the active orbitals. Quadrupole moments 
for configurations in the nuclei $^{150}$Dy, $^{148}$Dy, $^{150}$Gd, 
$^{148}$Gd, $^{144}$Gd, $^{143}$Eu and $^{142}$Sm 
were therefore calculated in three different ways:

\begin{itemize}
\item  By direct calculation at the appropriate equilibrium 
deformation; $Q_{exact}$ in Table~2.

\item  From the quadrupole moment of the superdeformed $^{152}$Dy yrast 
state and the sum of effective one-hole quadrupole moments;
$Q_{est}(q_{eff})$ in Table~2.

\item  From the quadrupole moment of the superdeformed $^{152}$Dy 
yrast state and the sum of effective quadrupole moments calculated
from single-particle quadrupole moments by the simple linear relations
eqs.\ (\ref{macp},\ref{macn}); $Q_{est}(q_{\nu})$ in Table~2.
\end{itemize}

\renewcommand{\arraystretch}{1.7}		

\begin{table}[htb]
\caption{Macroscopic quadrupole moments calculated in three different 
ways (see text) for SD configurations in selected nuclei with one or
several holes relative to the $^{152}$Dy reference nucleus.}
\label{tab2}\hspace{-0.1cm}
\par
\begin{tabular}{|clcccc|}
\hline
nucleus & configuration relative & $Q_{exact}$  & 
$Q_{est}(q_{eff})$ 
& $Q_{est}(q_{\nu})$  & $Q_{exact} - Q_{est}(q_{\nu})$\\ 
 & SD $^{152}$Dy yrast &  ($e$fm$^{2}$) & 
($e$fm$^{2}$) & ($e$fm$^{2}$) & ($e$fm$^{2}$)\\ 
\hline
$^{152}$Dy &  & 1893 &  &  &\\ 
$^{150}$Dy & $\nu 7^{-2}$ & 1725 & 1722 & 1715 & 10 \\ 
$^{150}$Gd & $\pi 6^{-2}$ & 1735 & 1733 & 1722 & 13 \\ 
$^{148}$Dy & $\nu 6^{-4}$ & 1701 & -- & 1714 & $-13$ \\ 
$^{148}$Gd & $\pi 6^{-2}\nu 7^{-2}$ & 1576 & 1562 & 1544 & 32 \\ 
$^{144}$Gd & $\pi 6^{-2}\nu 7^{-2}6^{-4}$ & 1381 & -- & 1365 & 16 \\ 
$^{143}$Eu & $\pi 6^{-3}\nu 7^{-2}6^{-4}$ & 1305 & -- & 1277 & 28 \\ 
$^{142}$Sm & $\pi 6^{-3}([541]1/2)^{-1}\nu 7^{-2}6^{-4}$ & 1232 & -- 
& 1198 & 34 \\ 
$^{142}$Sm & $\begin{array}{l} \pi 6^{-2}([541]1/2)^{-1}3^{-1} \\ 
\nu 7^{-1}6^{-4}([411]1/2)^{-1} \end{array}$  
& 1419 & -- & 1410 & 9 \\ \hline
\end{tabular}
\end{table}

When choosing the configurations, we have to make sure that if two
orbitals interact in the superdeformed region, both these orbitals should be 
either empty or occupied. This is the reason why the four $N=6$ neutron 
orbitals (2 of each signature) are treated as one entity. 

In Table~2, $Q_{est}(q_{eff})$ values are presented only for
configurations which have holes in the orbitals used to get the
effective one-hole quadrupole moments plotted in Fig.~4 while
$Q_{est}(q_{\nu})$ are calculated for all configurations.  The two
estimates based on $q_{eff}$ and $q_{\nu}$, respectively, differ by up
to 18 $e$fm$^{2}$ for four particles removed.  The difference between
the calculated value, $Q_{exact}$, and the single-particle estimate
$Q_{est}(q_{\nu})$ is at most 34 $e$fm$^{2}$. The corresponding
rms-value is 22 $e$fm$^{2}$ for all configurations in Table~2. It is
astonishing that the summation of effective quadrupole moments,
calculated from eqs.\ (\ref{macp}, \ref{macn}), describes the `exactly'
calculated values within 2\% for $^{142}$Sm, which is ten particles
away from $^{152}$Dy, and where the deformation has changed from
$\varepsilon =0.58$ to $\varepsilon =0.48$ and $\varepsilon =0.52$ for
the two studied configurations.

The two $^{142}$Sm configurations included have been measured in
experiment \cite{Hac98} and comparison with these data will be
discussed in subsection~3.5. In the Sm-bands, there are two orbitals
which have to be handled with extra care. Thus, due to the difference
in deformation the order of some orbitals closest to the Fermi-surface
are different in $^{152}$Dy and $^{142}$Sm.  At the deformation of
$^{152}$Dy, the proton $N_{osc}=5$ and neutron $N_{osc}=4$ closest to
the Fermi-surface are $\pi ([532]5/2)$ and $\nu ([404]9/2)$, i.e.\ the
corresponding points in Fig.~4 are constructed from configurations
with holes in these orbitals.  At the deformation of the Sm-bands on
the other hand, the $\pi ([541]1/2)$ and $\nu ([411]1/2)$ orbitals are
higher in energy and therefore, the bands are formed with holes in
these orbitals relative to the $^{152}$Dy bands. Consequently, their
single-particle moments should be used in the relations to get the
quadrupole moments in $^{142}$Sm, see Table~2. The orbitals are
relatively pure in the $^{142}$Sm configurations because, at the
relevant rotational frequencies, the crossings between the orbitals
occur at deformations somewhere between that of $^{152}$Dy and
$^{142}$Sm. Therefore, these configurations are anyway a good test on
how well additivity, based on $q_{\nu}$, works 10 particles away from
$^{152}$Dy.

\subsection{Microscopic polarization effects of one-particle states}

In a similar way as in section~3.1 we now calculate the microscopic
quadrupole moments $Q^{mic}=e\sum_{\nu =1}^{Z} q_{\nu}$ in
neighbouring nuclei to superdeformed $^{152}$Dy and plot $q_{eff}$
vs. $q_{\nu}$, see the solid lines in Fig.~6.
\begin{figure}[htb]
\psfig{figure=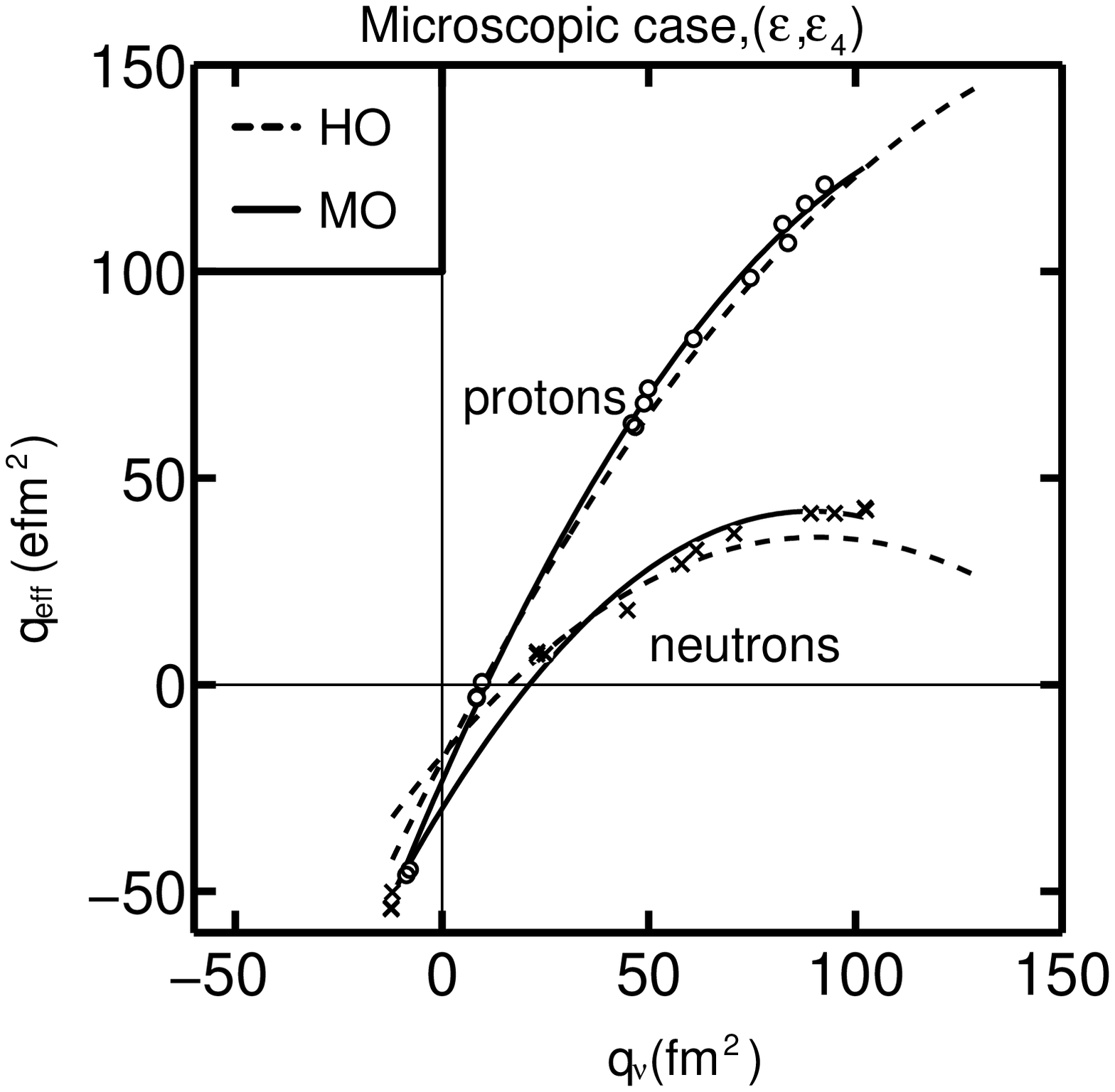,height=8cm}
\caption{Changes in the total microscopic quadrupole moment, $q_{eff}$, 
when a particle is added to or removed from the superdeformed core
plotted vs. the single-particle mass quadrupole moment, $q_{\nu}$.
The circles and crosses are obtained from calculations in the
Strutinsky-renormalized cranked MO with simultaneous
energy-minimization in the $\varepsilon$ and $\varepsilon_4$ directions
using $^{152}Dy$ as a core (values are given in Table~4). The solid
lines are quadratic least square fits to these points. The dashed
lines are obtained from analogous fits for the HO potential with the
($Z=60, N=80$) core, as defined in Fig.~7.}
\label{fig6}
\end{figure}  
Using the full MO, the relations no longer are approximatively linear
but rather quadratic,
\begin{equation}
q_{eff}=(-0.008q_{\nu}^2+2.28q_{\nu}-23.3)e{\rm fm}^2
\label{micpmo}
\end{equation}
for protons and 
\begin{equation}
q_{eff}=(-0.009q_{\nu}^2+1.61q_{\nu}-30.0)e{\rm fm}^2
\label{micnmo}
\end{equation}
for neutrons. This might suggest that it would be more proper to
express $q_{eff}$ not only as a function of the single-particle
quadrupole moment, but also of the hexadecapole moment. However, as
found below, the relations eqs.~(\ref{micpmo}) and (\ref{micnmo}) seem to
work well in the limited region of superdeformed nuclei with
$A=142-152$, so at present we will make no attempt to generalize eqs.\
(\ref{micpmo}) and (\ref{micnmo}).

In a similar way as in the macroscopic case, the reason why the microscopic 
relations are so different from what was 
found in the $Z=N$ HO calculations is now analyzed by
performing calculations where the different terms in the full MO 
relative to the HO are introduced one at a time.
 
First, allowing a different number of protons and neutrons and
calculating also for non-selfconsistent gaps give linear relations with changes
in the slopes in accordance with eq.~(\ref{scaling}). Introducing the
Strutinsky renormalization and cranking only has a minor effect on the
$q_{eff}$ vs. $q_{\nu}$ relations.

The result of including hexadecapole deformation $\varepsilon_4$ 
in the HO is shown in Fig.~7.
\begin{figure}[bt]
\psfig{figure=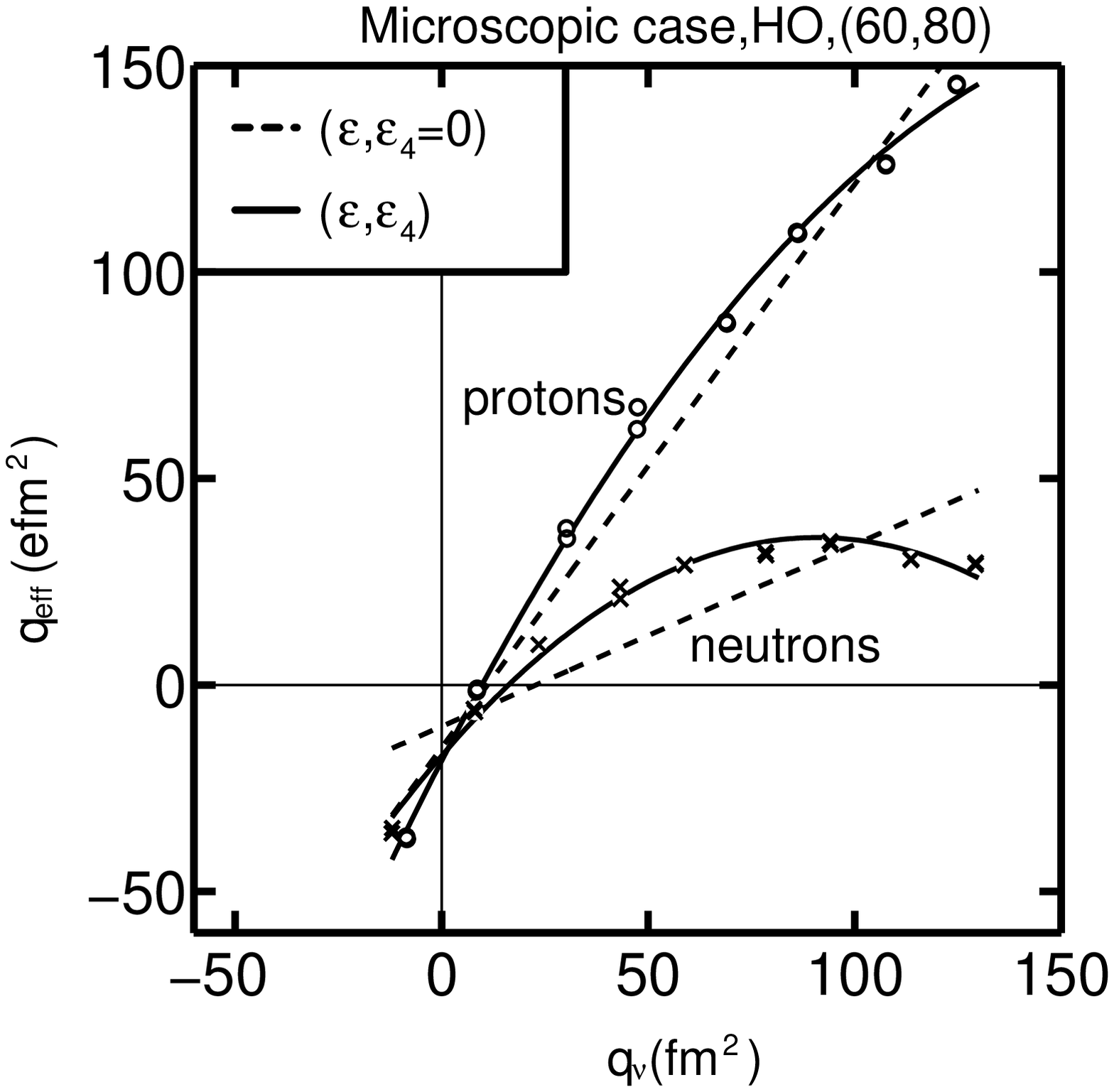,height=8cm}
\caption{Changes in the total microscopic quadrupole moment, $q_{eff}$, when
a particle is added to or removed from the $Z=60$ and $N=80$ core
plotted vs. the single-particle mass quadrupole moment, $q_{\nu}$.
The curves are least square fits to the calculations in
Strutinsky-renormalized cranked HO calculations. The dashed lines show the
relation when the energy is minimized only in $\varepsilon$ deformation
with $\varepsilon_4=0$. The solid lines (corresponding proton and neutron data
marked with circles and crosses, respectively) show the relation when
the energy is simultaneously minimized in the $\varepsilon$ and
$\varepsilon_4$ directions.}
\label{fig7}
\end{figure} 
As in the macroscopic case (see Fig.~5), the hexadecapole deformation
introduces a curvature, which is, however, more than twice as large in this
microscopic case. In Fig.~6 the MO result is compared to the HO
result, both including the hexadecapole deformation. The two curves
are seen to be rather similar, although the stronger polarization
effect for the MO than for the HO, discussed above for the macroscopic
case, can be seen also in the microscopic case. Furthermore, there is a small
increase of the curvature of the $q_{eff}$ vs.\ $q_{\nu}$ relation.

The reason why the introduction of the $\vec{l}\cdot \vec{s}$ and
$l^2$-terms rather increase the curvature in the microscopic case (Fig.~6) but
removes the curvature in the macroscopic case (Fig.~4) is not understood.

\subsection{Additivity of microscopic effective quadrupole moments}

The additivity in the microscopic case is checked, see Table~3, in the
same way as in the macroscopic case, i.e.\ the electric quadrupole
moment is calculated in three different ways $Q_{exact}$,
$Q_{est}(q_{eff})$, and $Q_{est}(q_{\nu})$, as described in
subsection~3.2.

The result of this test is that the additivity seems to work 
with a similar accuracy in this model as in the macroscopic approach.
This is illustrated in Fig.~8 where 
\begin{figure}[htb]
\psfig{figure=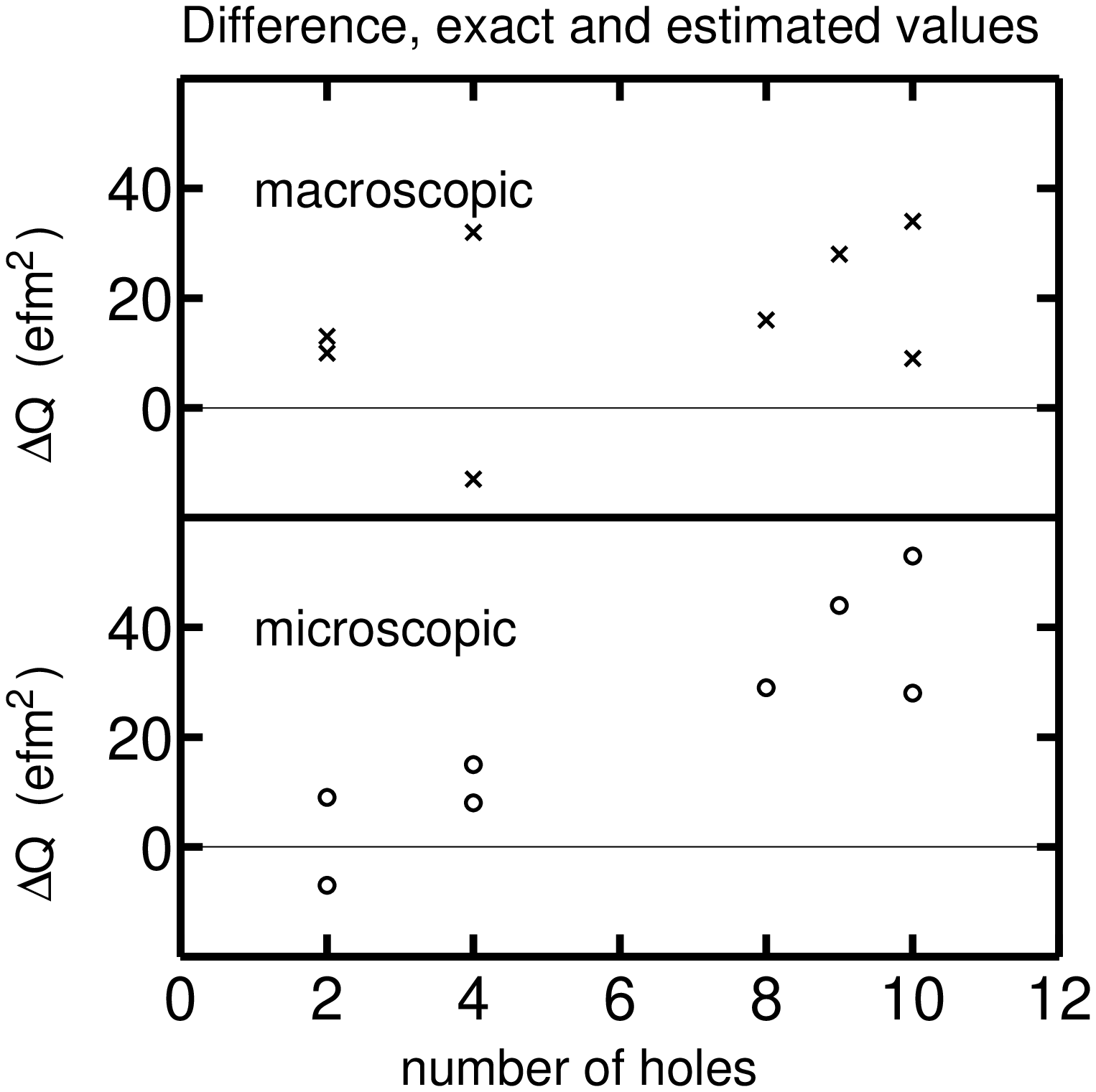,height=8cm}
\caption{The difference between the calculated and estimated 
(based on $q_{\nu}$) quadrupole moment as a function of number of
holes relative to the superdeformed $^{152}$Dy core. The upper panel
shows the macroscopic result from Table~2 while the lower panel shows
the corresponding microscopic result from Table~3.}
\label{fig8}
\end{figure}
the difference between the calculated, $Q_{exact}$, and the single
particle estimated quadrupole moments, $Q_{est}(q_{\nu})$, is plotted
as a function of number of holes relative to the superdeformed
$^{152}$Dy core.
 
For the three configurations where the quadrupole moment has been
estimated based on both $q_{eff}$ and $q_{\nu}$, the maximum deviation
is 7 $e$fm$^{2}$, see Table~3. The rms value for
$Q_{exact}-Q_{est}(q_{\nu})$ is 29 $e$fm$^{2}$ for the eight
configurations considered in Table~3, while the maximum deviation is
53 $e$fm$^{2}$. In the microscopic case, summing effective quadrupole
moments, calculated from eqs.~(\ref{micpmo}, \ref{micnmo}), describes
the two $^{142}$Sm configurations with a 4 \% 
accuracy.

\renewcommand{\arraystretch}{1.7}		

\begin{table}[htb]
\caption{Microscopic quadrupole moments calculated for SD configurations 
in selected nuclei with one or several holes relative to the $^{152}$Dy 
reference nucleus.}
\label{tab3}\hspace{-0.1cm}
\par
\begin{tabular}{|clcccc|}
\hline
nucleus & configuration relative & $Q_{exact}$  & 
$Q_{est}(q_{eff})$ 
& $Q_{est}(q_{\nu})$  & $Q_{exact} - Q_{est}(q_{\nu})$\\ 
 & SD $^{152}$Dy yrast &  ($e$fm$^{2}$) & 
($e$fm$^{2}$) & ($e$fm$^{2}$) & ($e$fm$^{2}$)\\ 
\hline
$^{152}$Dy &  & 1810 &  &  & \\ 
$^{150}$Dy & $\nu 7^{-2}$ & 1722  & 1725 & 1729  & $-7$ \\ 
$^{150}$Gd & $\pi 6^{-2}$ & 1607 & 1605 &  1598 & 9 \\ 
$^{148}$Dy & $\nu 6^{-4}$ & 1675 & -- & 1660 & 15 \\ 
$^{148}$Gd & $\pi 6^{-2}\nu 7^{-2}$ & 1528 & 1520 & 1517  & 8 \\ 
$^{144}$Gd & $\pi 6^{-2}\nu 7^{-2}6^{-4}$ & 1396 & -- & 1367  & 29\\ 
$^{143}$Eu & $\pi 6^{-3}\nu 7^{-2}6^{-4}$ & 1303 & -- & 1259  & 44\\ 
$^{142}$Sm & $\pi 6^{-3}([541]1/2)^{-1}\nu 7^{-2}6^{-4}$ & 1213 & -- & 1160  
& 53\\ 
$^{142}$Sm & $\begin{array}{l} \pi 6^{-2}([541]1/2)^{-1}3^{-1} \\ 
\nu 7^{-1}6^{-4}([411]1/2)^{-1} \end{array}$  
& 1395 & -- & 1367 & 28 \\ \hline
\end{tabular}
\end{table}

\subsection{Comparison between macroscopic, microscopic and experimental 
effective quadrupole moments} 

The two models studied give quite simple relations (eqs.~(\ref{macp},
\ref{macn}) and eqs.~(\ref{micpmo}, \ref{micnmo})) between $q_{eff}$ and
$q_{v}$, but their predictions for specific orbitals are rather
different as can be seen in Fig.~9. 
\begin{figure}[htb]
\psfig{figure=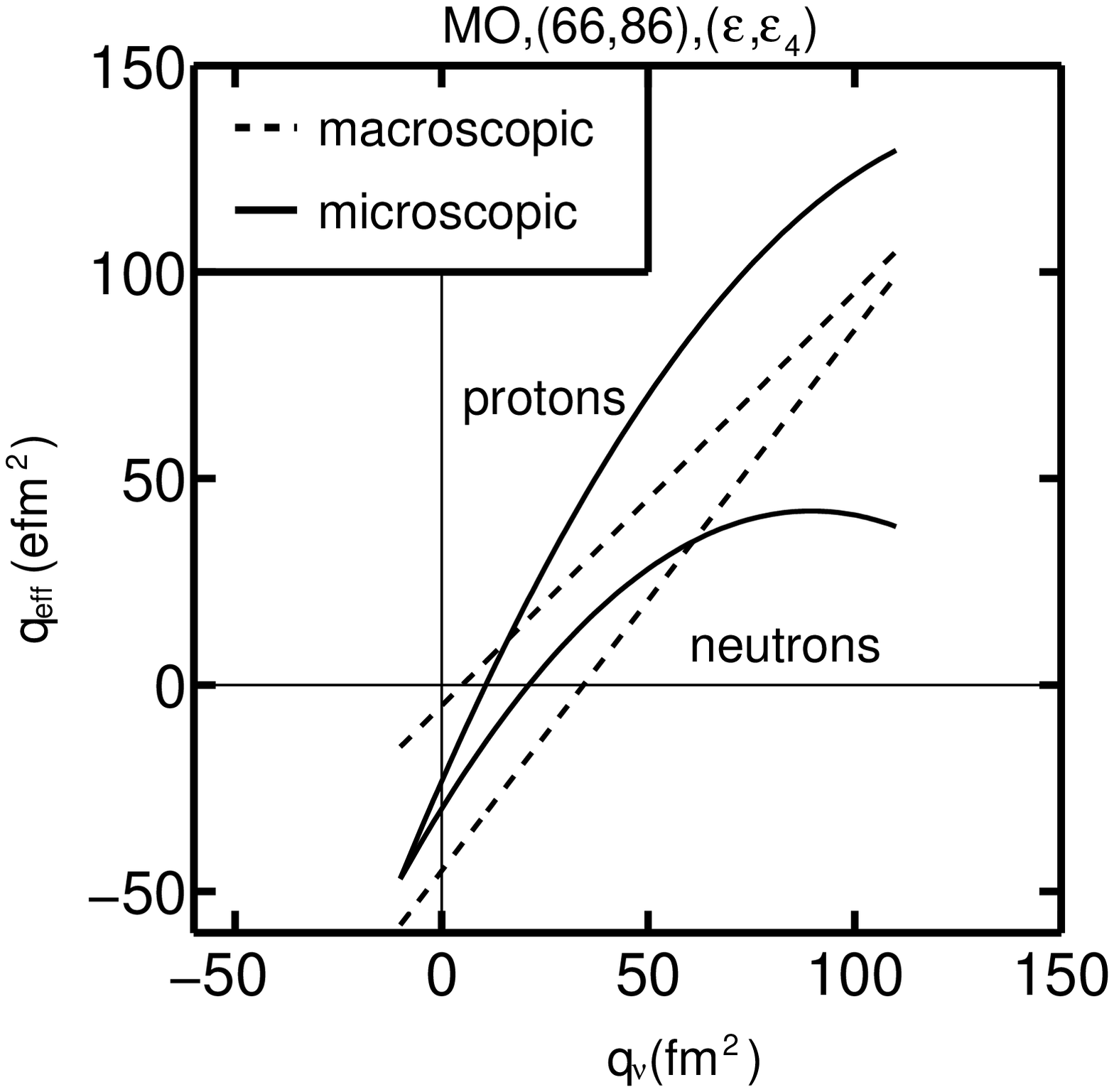,height=8cm}
\caption{Effective electric quadrupole moments versus 
single-particle mass quadrupole moments calculated in the MO for one
particle/hole outside the superdeformed $^{152}$Dy core. Dashed lines are
used for the macroscopic and solid lines for the microscopic
method. It is clear that for specific combinations of protons and
neutrons removed from the $^{152}$Dy core, the two methods could lead
to rather different results.}
\label{fig9}
\end{figure}

In the macroscopic relations neutrons and protons with high
$q_{\nu}$-value have similar effects, while the microscopic relations
give very different values for protons and neutrons, with exception
for the lowest $q_{\nu}$-values.  In the microscopic case for neutrons
the maximum $q_{eff}$ is not obtained from the maximum $q_{\nu}$. In
Table~4 $q_{\nu}$ and $q_{eff}$-values are given for several different
orbitals close to the Fermi-surface in $^{152}$Dy.  The
$q_{eff}$-values are shown both for the macroscopic and the
microscopic case. For comparison values from Skyrme-Hartree-Fock
calculations are also presented. By using these effective quadrupole
moments together with eq.~(\ref{Qest}) (with the scalings discussed
below) electric quadrupole moments can be estimated for a large number
of superdeformed configurations in a rather large region around
$^{152}$Dy. There are also configurations with only $q_{\nu}$-values
given. From those one can estimate effective electric quadrupole
moments by using eqs.~(\ref{macp},\ref{macn}) and
eqs.~(\ref{micpmo},\ref{micnmo}), and through eq.~(\ref{Qest}) get
good estimates of electric quadrupole moments for many more
configurations in this region.

\renewcommand{\arraystretch}{1.2}
\begin{table}[htb]
\caption{   }
\setbox1=
\vbox{
Effective quadrupole moments calculated for orbitals in the 
SD $A=150$ region. The macroscopic and microscopic values, calculated
in $^{152}$Dy $\pm$ 1 particle, are the ones used when producing
Fig.~4 and Fig.~6. For other orbitals only $q_{\nu}$-values is given. They
can be used together with the relations eqs.~(\ref{macp},
\ref{macn}) and eqs.~(\ref{micpmo}, \ref{micnmo}) to estimate macroscopic and 
microscopic effective quadrupole moments. The Skyrme-Hartree-Fock
calculations are from ref.~\cite{Sat96}.}
\label{tab4}\hspace{-0.1cm}
\par \scriptsize
\setbox2=
\hbox{
\begin{tabular}{|cccccc|cccccc|}
\hline
orbital & $q_{\nu}$ & $q^{mac}_{eff}$ & $q^{mic}_{eff}$ 
& $q^{SkP}_{eff}$ & $q^{SkM^{*}}_{eff}$ &
orbital & $q_{\nu}$ & $q^{mac}_{eff}$ & $q^{mic}_{eff}$ 
& $q^{SkP}_{eff}$ & $q^{SkM^{*}}_{eff}$  \\ 
& ($e$fm$^{2}$) & ($e$fm$^{2}$) & ($e$fm$^{2}$) & ($e$fm$^{2}$) & 
($e$fm$^{2}$) & 
& ($e$fm$^{2}$) & ($e$fm$^{2}$) & ($e$fm$^{2}$) & ($e$fm$^{2}$) & 
($e$fm$^{2}$)  \\ \hline
\multicolumn{6}{|c|}{proton holes} & 
\multicolumn{6}{c|}{neutron holes} \\
$\pi ([301]1/2^-)^{-1}$ & -7.8 & -5.7 &  -44.8 & -15 & -18 & 
$\nu ([404]9/2^-)^{-1}$ & -12.3  & -66.5 &  -54.0 & & \\  
$\pi ([301]1/2^+)^{-1}$ & -8.6& -5.5 &  -46.1 & -18 & -16 & 
$\nu ([404]9/2^+)^{-1}$ & -12.3 & -66.8 &  -54.3 & & \\ 
$\pi ([413]7/2^-)^{-1}$ & 8.3 &  13.6 &   -3.0 & & & 
$\nu ([523]7/2^-)^{-1}$ & 23.0 & -8.5 &    7.9 & & \\ 
$\pi ([413]7/2^+)^{-1}$ & 8.3 &  13.0 &   -3.3 & & & 
$\nu ([523]7/2^+)^{-1}$ & 23.0 & -9.3 &    7.3 & & \\ 
$\pi ([532]5/2^-)^{-1}$ & 46.7 &  54.7 &   63.2 & & & 
$\nu ([642]5/2^-)^{-1 (a)}$ & 58.0 &  32.6 &   29.2 &  22 &  22 \\ 
$\pi ([532]5/2^+)^{-1}$ & 45.9 &  52.3 &   62.4 & & & 
$\nu ([642]5/2^+)^{-1 (a)}$ & 70.6 &  61.7 &   36.6 &  24 &  24 \\ 
$\pi ([651]3/2^-)^{-1}$ & 83.7 &  81.8 &  106.9 &  96 &  96 & 
$\nu ([770]1/2^-)^{-1}$ & 102.3 &  88.3 &   42.7 &  59 &  48 \\ 
$\pi ([651]3/2^+)^{-1}$ & 74.6 &  78.5 &   98.5 &  89 &  88 & 
$\nu ([770]1/2^+)^{-1}$ & 102.6 &  82.6 &   42.3 &  57 &  48 \\ 
$\pi ([541]1/2^-)^{-1}$ & 71.6 &   &    &   &   & 
$\nu ([411]1/2^-)^{-1}$ & 8.7 &   & &  18  & 16    \\ 
$\pi ([541]1/2^+)^{-1 }$ & 68.3 &   &    &   &   & 
$\nu ([411]1/2^+)^{-1}$ & 9.4 &   & & 15  & 13    \\ 
$\pi ([660]1/2^-)^{-1}$ & 81.4 &   &    &   &   & 
$\nu ([651]1/2^-)^{-1 (a)}$ & 78.5 &   & & 43  & 28    \\
$\pi ([660]1/2^+)^{-1}$ & 84.6 &   &    &   &   & 
$\nu ([651]1/2^+)^{-1 (a)}$ & 66.1 &   & & 43   & 30    \\   \hline
\multicolumn{6}{|c|}{proton particles} & 
\multicolumn{6}{c|}{neutron particles} \\
$\pi ([413]5/2^-)$ & 9.6 &  16.7 &    0.6 & & & 
$\nu ([402]5/2^-)$ & -12.0 & -60.4 &  -50.2 & -44 & -38 \\  
$\pi ([413]5/2^+)$ & 9.6 &  16.7 &    10.6 & & & 
$\nu ([402]5/2^+)$ & -12.0 & -60.4 &  -50.2 & -44 & -38 \\ 
$\pi ([532]3/2^-)$ & 48.8 &  61.5 &   70.2 & & & 
$\nu ([521]3/2^-)$ & 24.9 & -5.0 &    7.4 &   0 &  -1 \\ 
$\pi ([532]3/2^+)$ & 49.9 &  60.5 &   71.7 & & & 
$\nu ([521]3/2^+)$ & 24.9 & -5.0 &    7.4 &   0 &  -1 \\ 
$\pi ([642]5/2^-)$ & 82.4 &  92.9 &  111.5 & & & 
$\nu ([640]1/2^-)$ & 44.8 &  5.5 &   18.0 & & \\ 
$\pi ([642]5/2^+)$ & 60.8 &  64.2 &   83.7 & & & 
$\nu ([640]1/2^+)$ & 61.5 &  40.6 &   32.6 & & \\ 
$\pi ([770]1/2^-)$ & 92.6 &  101.1 &  121.0 & & & 
$\nu ([761]3/2^-)$ & 89.2 &  68.4 &   41.5 &  46 &  41 \\  
$\pi ([770]1/2^+)$ & 87.9 &  94.1 &  116.4 & & & 
$\nu ([761]3/2^+)$ & 95.0 &  72.3 &   41.5 &  41 &  28 \\ \hline
\end{tabular}}
\normalsize
\setbox3=
\vbox{
(a) The orbitals $\nu ([642]5/2)$ and $\nu ([651]1/2)$ are mixed in
the Dy-region and are not pure in $^{152}$Dy. If none or both orbitals
of the same parity is unoccupied the quadrupole moment can be correct
calculated from these values, else extra care should be taken. The
labels are valid for the Skyrme-Hartree-Fock values.}
\centerline{\rotl{1} \rotl{2} \rotl{3}}
\end{table}

\begin{table}[htb]
\caption{Macroscopic and microscopic quadrupole moments are compared with 
experiments. The holes are specified relative to the reference, 
superdeformed $^{152}$Dy yrast configuration.}
\label{tab5}\hspace{-0.1cm}
\par
\begin{tabular}{|clccccc|}
\hline
nucleus & configuration relative & $Q^{exp}$ & 
$0.92 Q^{mac}$ &  $0.96 Q^{mic}$  & $Q^{exp}-0.92 Q^{mac}$ & 
$Q^{exp}-0.96 Q^{mic}$\\ 
 & SD $^{152}$Dy yrast &  ($e$b$^{2}$) &  ($e$b$^{2}$) & ($e$b$^{2}$) & 
($e$b$^{2}$) &  ($e$b$^{2}$) \\ 
\hline
$^{152}$Dy & & 17.5$^{a)}$ & 17.4  & 17.4 & 0.1 & 0.1\\
$^{151}$Dy & $\nu 7^{-1}$ & 16.9$^{b)}$ & 16.6 & 17.0 & 0.3 & $-0.1$\\
$^{151}$Tb & $\pi 6^{-1}$ & 16.8$^{a)}$ & 16.6  & 16.3 & 0.2  & 0.5 \\
$^{149}$Gd & $\pi 6^{-2}\nu 7^{-1}$ & 15.0$^{a)}$ & 15.2 &  15.1 & $-0.2$ 
& $-0.1$\\ 
$^{149}$Gd & $\pi 6^{-2}\nu 6^{-1}$ & 15.6$^{a)}$ & 15.4  & 15.1 & 0.2 & 0.5\\ 
$^{149}$Gd & $\nu 6^{-1}3^{-1} \nu 7^{-1}$ & 15.2$^{a)}$ & 16.0 & 16.2 
& $-0.8$ & $-1.0$\\ 
$^{149}$Gd$^{d)}$ & $\pi 3^{-2}\nu 4^{-1}$ & 17.5$^{a)}$ & 17.7 &  18.3 
& 0.2 & 
$-0.8$ \\ 
$^{148}$Gd & $\pi 6^{-2}\nu 7^{-1}6^{-1}$ & 14.6$^{a)}$ & 14.7  & 14.7 
& $-0.1$ & $-0.1$ \\ 
$^{148}$Gd & $\pi 6^{-2}\nu 7^{-1}6^{-1}$ & 14.8$^{a)}$ & 14.7 &  14.7 
& 0.1 & 0.1 \\ 
$^{148}$Gd$^{d)}$ & $\pi 3^{-2}\nu 4^{-2}$ & 17.8$^{a)}$ & 18.0 &  18.5 & 
$-0.2$ & $-0.7$ \\ 
$^{142}$Sm & $\pi 6^{-3}5^{-1}\nu 7^{-2}6^{-4}$ & 11.7$^{c)}$ & 11.3  & 11.6 
& 0.4 & 0.1\\ 
$^{142}$Sm$^{e)}$ & $\pi 6^{-2}5^{-1}3^{-1}\nu 7^{-1}6^{-4}4^{-1}$ 
& 13.2$^{c)}$ & 13.1  & 13.4 & 0.1 & $-0.2$ \\ 
\hline
\end{tabular}
The experimental data are from $^{a)}$ ref.~\cite{Sav96}, $^{b)}$ ref.\
\cite{Nis97}, $^{c)}$ ref.~\cite{Hac98}. In the configurations marked
$^{d)}$ the hole is forced to the $\nu ([411]1/2)^{-1}$ orbit in order not
to mix with $\nu ([404]9/2)^{-1}$ orbit while in the configuration
marked $^{e)}$ the hole naturally comes in the $\nu ([411]1/2)^{-1}$ orbit.
\end{table} 
In the comparison with experiments, see Table~5, the calculated
quadrupole moments are scaled with factors to give approximately the
same value for $^{152}$Dy as the experimental data. This is partly
motivated by the uncertainties in the absolute values obtained in
experiment, due to the uncertainties in stopping powers. Also in
Woods-Saxon, Hartree-Fock with Skyrme force, and cranked relativistic
mean field calculations \cite{Naz89,Sat96,Afa96} the values are
systematically higher than in experiment. We see that the macroscopic
method reproduces the relative changes with somewhat better accuracy
in this region. The rms-values between the experimental and scaled
theoretical quadrupole moments are 31 $e$fm$^2$ and 48 $e$fm$^2$ for
macroscopic and microscopic models, respectively. The configuration
with the largest discrepancy, $\nu 6^{-1}3^{-1} \nu 7^{-1}$ in
$^{149}$Gd, has the largest discrepancy also in the
Skyrme-Hartree-Fock calculation \cite{Sat96} and the error is
almost the same.

It is also interesting to note that the contribution to the change of
quadrupole moments coming from protons and neutrons are very different
in the two approaches. In the macroscopic model 48 \% 
of the total change in quadrupole moment from the superdeformed yrast 
band in$^{152}$Dy to the exited band ($\pi6^{-3}5^{-1}\nu 7^{-2}6^{-4}$) 
in $^{142}$Sm comes from adding $q_{eff}$ for the removed protons,
while the corresponding number is 64 \% 
in the microscopic model. 

In the mean field calculation based on e.g.\ the modified oscillator
or Woods-Saxon potential, it is required that the proton and neutron
deformation are exactly the same, namely the proton and neutron
single-particle potentials are defined for the same deformation
parameters. Then in the macroscopic method to calculate quadrupole
moments, also the matter distribution is assumed to have the same
deformation. As discussed in subsect.~2.2 for the HO this explains why
for a $Z=N$ nucleus $\left( b^{mac}\right) _p =\left(
b^{mac}\right) _n \approx 1$, while the microscopic approach leads to
different deformations leading to $\left( b^{mic}\right) _p =1.5$
and $\left( b^{mac}\right) _n=0.5$. Then, as seen in Fig.~9, these
expressions are modified by different factors but general features are
still the same in Nilsson-Strutinsky-cranking MO calculations for
$Z<N$ nuclei. The numbers in Table~5 do rather support the macroscopic
formula, e.g.\ when comparing $^{152}$Dy and $^{149}$Gd configurations
with configurations in $^{151}$Dy and $^{148}$Gd which differ by one
$N=7$ neutron. This $N=7$ neutron appears to have a large influence on
the measured quadrupole moments in somewhat closer agreement with the
macroscopic than the microscopic calculations. On the other hand,
measured quadrupole moments in $^{131,132}$Ce \cite{Cla96} indicate a
very small polarization for an $N=6$ neutron in this region, even
smaller than suggested by our microscopic calculations. We can conclude that
more experimental data with high accuracy combined with comparison
with selfconsistent calculations are required to disentangle the
polarization properties of protons and neutrons, respectively.

\section{Summary}
We have investigated the polarization effects of a particle on a
well-deformed core in the harmonic oscillator (HO) potential as well
as in the modified oscillator (MO) potential. Two different ways to
calculate the quadrupole moment (and thereby the polarization effect)
were considered. In the microscopic approach the electric
single-particle quadrupole moments are summed at the appropriate
equilibrium deformations, while in the macroscopic approach the
quadrupole moment is calculated by considering the nuclear charge as
uniformly distributed over its volume, again at the appropriate
equilibrium deformation.  Averaging over protons and neutrons, the two
models were found to give similar results even though the individual
proton and neutron contributions turned out to be rather different.

In the pure HO model, it was found for a $Z=N$ system, that the change
of the electric quadrupole moment when a particle (or hole) is added,
$q_{eff}$, can be described by a simple linear relation in the
single-particle mass quadrupole moment, $q_{\nu}$:
$q_{eff}=e(bq_{\nu}+a)$. Analytical expressions were derived for the
deformation and mass dependence of the parameters $a$ and $b$.  It
turned out that in the microscopic model, $b=1.5$ for protons and
$b=0.5$ for neutrons while in the macroscopic model, $b$ showed some
variation but was close to one for all deformations and particle
numbers, for both protons and neutrons. These differences were
explained from the way the proton and neutron matter distributions are
assumed to adjust to each other when the equilibrium deformation of
the individual systems are different. Allowing $Z \neq N$, neutron
excess led to a decrease of the $b$-values, especially for protons.

In the macroscopic case, $a$ was essentially equal to zero for
protons. In the other three cases, it was negative for prolate shapes
and positive for oblate shapes, and scales with mass $A$ approximately
in the same way as the single-particle quadrupole moment, i.e.\
proportional to $A^{2/3}$.  The fact that the parameter $a$ is
positive for oblate shapes and negative for prolate shapes is easily
understood. The quadrupole moment of the added particle must overcome
some value in order to increase the core deformation, and this value
is obviously positive for prolate shapes, negative for oblate shapes
and zero for spherical shapes.
  
In the Nilsson-Strutinsky cranking calculations we used the MO
potential as the microscopic potential, and calculated effective
quadrupole moments around the superdeformed core of $^{152}$Dy.  Both
the macroscopic approach and the microscopic approach were used. From
a basic point of view the microscopic way of calculating quadrupole
moments appears most reasonable. On the other hand the macroscopic
approach has been used frequently in previous realistic calculations
and been found to work well. In the macroscopic model numerical
calculation indicated that the linear relation between $q_{eff}$ and
$q_{\nu}$ were valid and the polarization factors were then numerically
obtained as $b_p=1.02$, $b_n=1.30$, $a_p=5.4$ fm$^{2}$ and $a_n=-43.7$
fm$^{2}$.  These numbers were different from the factors deduced from
the pure HO in the same macroscopic way for an $Z=66, N=86$
superdeformed nucleus ($b_p=0.78$ and $b_n=0.93$).  The reason for
this deviation was explained as being due to the decreased stiffness
of the potential energy surface around the minimum for the MO. In the
microscopic model there appeared to be a stronger dependence on
hexadecapole deformation which led to a need for quadratic
relations.

Additivity of effective quadrupole moments in superdeformed nuclei was
investigated and found to work surprisingly well. 
Adding $q_{eff}$-values, calculated from one-hole and one-particle states
outside a superdeformed core of $^{152}$Dy, quadrupole moments could
be well described in an extensive region of superdeformed nuclei.
Similar results using the Skyrme-Hartree-Fock method were previously 
obtained by Satu{\l}a {\it et al}.\ \cite{Sat96}.
Furthermore, using the simple relations for $q_{eff}$ as a function of
$q_{\nu}$, quadrupole moments could be estimated in an even larger
region using only the total quadrupole moment of the core $^{152}$Dy
together with $q_{\nu}$-values for the active single-particle
orbitals as input. For example, in the macroscopic case, the 10-hole
configurations describing two observed superdeformed band in $^{142}$Sm 
were both estimated within a 2 \% 
accuracy relative to the values obtained from a full calculation for
these bands. In the microscopic case the additivity worked with a 
somewhat smaller accuracy and we obtained a 4 \% 
accuracy for the two $^{142}$Sm bands.

From the (bare) single-particle quadrupole moments given in Table~4 it
should be possible to estimate total electric quadrupole moments with
a reasonably accuracy for configurations in a quite extended region of
superdeformed $A \sim 150$ nuclei.

The experimental data are reproduced in a good way by the theoretical
calculations with somewhat smaller discrepancies using the macroscopic 
method, see also refs.~\cite{Sav96,Nis97,Hac98}.

The surprising accuracy of the additivity suggests the possibility of
a shell-model type description of superdeformed nuclei, utilizing a
superdeformed core and a valence space consisting of superdeformed
one-particle (one-hole) states.

\vspace{1cm}

\noindent We are grateful to A.V.\ Afanasjev for useful comments on this
manuscript. I.R.\ and S.{\AA}.\ thank the Swedish National Research
Council (NFR) for financial support.


\newpage

\newpage

\newpage

\begin{thebibliography}{99}

\bibitem{Twi86} P.J.~Twin, B.M.~Nyak\'{o}, A.H.~Nelson, J.~Simpson, 
M.A.~Bentley, H.W.~Cranmer-Gordon, P.D.~Forsyth, D.~Howe, A.R.~Mokhtar, 
J.D.~Morrison, J.F.~Sharpey-Schafer, and G.~Sletten, Phys.\ Rev.\ Lett.\ 
{\bf 57} (1986) 811.

\bibitem{Ben87} M.A.~Bentley, G.C.~Ball, H.W.~Cranmer-Gordon,  P.D.~Forsyth, 
D.~Howe, A.R.~Mokhtar, J.D.~Morrison, J.F.~Sharpey-Schafer, P.J.~Twin,
B.~Fant, C.A.~Kalfas, A.H.~Nelson, J.~Simpson, and G.~Sletten,
 Phys.\ Rev.\ Lett.\ {\bf 59} (1987) 2141.

\bibitem{Sav96} H.~Savajols, A.~Korichi, D.~Ward, D.~Appelbe, 
G.C.~Ball, C.~Beausang, F.A.~Beck, T.~Byrski, 
D.~Curien, P.~Dagnall,  G.~de France, 
D.~Disdier, G.~Duch\^ene, S.~Erturk, C.~Finck, S.~Flibotte, 
B.~Gall, A.~Galindo-Uribarri, B.~Haas, G.~Hackman, 
V.P.~Janzen, B.~Kharraja, J.C.~Lisle, J.C.~Merdinger, 
S.M.~Mullins, S.~Pilotte, D.~Pr\'evost, D.C.~Radford, 
V.~Rauch, C.~Rigollet, D.~Smalley, M.B.~Smith, O.~Stezowski, 
J.~Styczen, Ch.~Theisen, P.J.~Twin, J.P.~Vivien, 
J.C.~Waddington, K.~Zuber, and I.~Ragnarsson, 
 Phys.\ Rev.\ Lett.\ {\bf 76} (1996) 4480.

\bibitem{Nis97} D.~Nisius, R.V.F.~Janssens, E.F.~Moore,
P.~Fallon, B.~Crowell, T.~Lauritsen, G.~Hackman,
I.~Ahmad, H.~Amro, S.~Asztalos, M.P.~Carpenter,
P.~Chowdhury, R.~M.~Clark, P.J.~Daly, 
M.A.~Deleplanque, R.M.~Diamond, S.M.~Fischer,
Z.W.~Grabowski, T.L.~Khoo, I.Y.~Lee, A.O.~Macchiavelli, 
R.H.~Mayer, F.S.~Stephens, A.V.~Afanasjev,
and  I.~Ragnarsson, Phys.\ Lett.\  {\bf B392} (1997) 18.


\bibitem{Hac98} G.~Hackman, R.V.F.~Janssens, E.F.~Moore, D.~Nisius,
I.~Ahmad, M.P.~Carpenter, S.M.~Fischer,  T.L.~Khoo, T.~Lauritsen, and 
P.~Reiter, Phys.\ Lett.\ {\bf B416} (1998) 268.

\bibitem{Mot58} B.R.~Mottelson, Cours de l'Ecole d'Et\'{e} de 
Physique Th\'{e}orique des Houches 1958, Dunod (1959) 283, 
also B.R.\ Mottelson, Nordita publications No.~20

\bibitem{Afa96} A.V.~Afanasjev, J.~K\"{o}nig, and P.~Ring, Nucl.\ Phys.\
{\bf A608} (1996) 107

\bibitem{Ben88} T.~Bengtsson, I.~Ragnarsson, and S.~{\AA}berg, 
Phys.\ Lett.\ {\bf B208}  (1988) 39.

\bibitem{Del89} M.A.~Deleplanque, C.W.~Beausang, J.~Burde, R.M.~Diamond, 
F.S.~Stephens, R.J.~McDonald, and J.E.~Draper, Phys.\ Rev.\ {\bf C39},
(1989) 1651.

\bibitem{Rag91} I.\ Ragnarsson, Phys.\ Lett.\ {\bf B264} (1991) 5.

\bibitem{Rag93} I.\ Ragnarsson, Nucl.\ Phys.\ {\bf A557} (1993) 167c.

\bibitem{preprint} A.V.\ Afanasjev, G.A.~Lalazissis, and P.~Ring,
to be publ., see also lanl nucl-th/9801038 


\bibitem{Sat96} W.\ Satu{\l}a, J.~Dobaczewski, J.~Dudek, and W.~Nazarewicz, 
Phys.\ Rev.\ Lett.\ {\bf 77} (1996) 5182.

\bibitem{BM2} A.~Bohr and B.R.~Mottelson, Nuclear Structure, vol.\ II,
W.A.~Benjamin Inc.\ (1975)


\bibitem{Nil55} S.G.\ Nilsson, Mat.\ Fys.\ Medd.\ Dan.\ Vid.\ Selsk.\ 
{\bf 29} (1955) no 16.

\bibitem{Lar72} S.E.\ Larsson, Phys.\ Scripta {\bf 8} (1973) 17.

\bibitem{Cer79} M.\ Cerkaski and  Z.~Szyma\'{n}ski, Acta Phys.\ Polon.\ 
{\bf B10} (1979) 163.

\bibitem{Nil95} S.G.\ Nilsson and I.~Ragnarsson, Shapes and Shells in 
Nuclear Structure, Cambridge University Press (1995).

\bibitem{Sak89} H. Sakamoto and T. Kishimoto,Nucl.\ Phys.\
{\bf A501} (1989) 205 


\bibitem{Nil69} S.G.\ Nilsson, C.F.~Tsang, A.~Sobiczewski, Z~Szyma\'{n}ski,
S.~Wycech, C.~Gustafsson, I.-L.~Lamm, P.~ M\"{o}ller, and B.~Nilsson, 
Nucl.\ Phys.\ {\bf A131} (1969) 1.


\bibitem{Haa93} B.\ Haas, V.P.~Janzen, D.~Ward, H.R.~Andrews, D.C.~Radford,
D.~Pr\'{e}vost, J.A.~Kuehner A.~Omar, J.C.~Waddington, T.E.~Drake, 
A.~Galindo-Uribarri, G.Zwartz, S.~Flibotte, P.~Taras, and I.~Ragnarsson, 
Nucl.\ Phys.\ {\bf A561} (1993) 251.

\bibitem{Ben85} T.\ Bengtsson and I.\ Ragnarsson, Nucl.\ Phys.\ {\bf A436}
(1985) 14.

\bibitem{Naz89} W.\ Nazarewicz, R.~Wyss and A.~Johnson, Nucl.\ Phys.\ 
{\bf A503} (1989) 285.

\bibitem{Rag80} I.\ Ragnarsson, T.~Bengtsson, G.~Leander, and S.~{\AA}berg, 
Nucl.\ Phys.\ {\bf A347} (1980) 287.

\bibitem{Cla96} R.M.~Clark, I.Y.~Lee, P.~Fallon, D.T.~Joss, S.J.~Asztalos,
J.A.~Becker, L.~Bernstein, B.~Cederwall, M.A.~Deleplanque,  R.M.~Diamond, 
L.P.~Farris, K.~Hauschild, W.H.~Kelly, A.O.~Macchiavelli, 
P.J.~Nolan, N.~O'Brien, A.T.~Semple, F.S.~Stephens, and R.~Wadsworth,
Phys.\ Rev.\ Lett.\ {\bf 76} (1996) 3510.

\end{thebibliography}
\end{document}